\documentclass[11pt,preprint]{aastex}
\usepackage{graphicx}
\usepackage{epstopdf}
\usepackage{subfigure}

\shortauthors{Berg et al.}
\title{Re-examining High Abundance SDSS Mass-Metallicity Outliers: High N/O, Evolved Wolf-Rayet Galaxies? }

\author{Danielle A. Berg\altaffilmark{1}, Evan D. Skillman\altaffilmark{1}, Andrew R. Marble\altaffilmark{2,3}}

\altaffiltext{1}{Department of Astronomy, University of Minnesota, 116 Church St. SE, Minneapolis, MN 55455; berg@astro.umn.edu; skillman@astro.umn.edu}
\altaffiltext{2}{Steward Observatory, University of Arizona, 933 N Cherry Ave, Tucson, AZ 85721}
\altaffiltext{3}{National Solar Observatory, 950 N Cherry Ave, Tucson, AZ 85719; amarble@nso.edu} 

\begin{document}
\begin{abstract}
We present new MMT spectroscopic observations of four dwarf galaxies
representative of a larger sample observed by the Sloan Digital Sky
Survey (SDSS) and identified by \citet{peeples08} as low-mass, high
oxygen abundance outliers from the mass-metallicity relation.
\cite{peeples08} showed that these four objects (with metallicity estimates
of $8.5\leq12+\log{(\mbox{O/H})}\leq8.8$) have oxygen abundance offsets
of 0.4-0.6 dex from the M$_B$ luminosity-metallicity relation.
Our new observations extend the wavelength coverage to include the
[\ion{O}{2}] $\lambda\lambda$3726,3729 doublet, which adds leverage in
oxygen abundance estimates and allows measurements of N/O ratios.
All four spectra are low excitation, with relatively high N/O ratios
($\mbox{N/O}\gtrsim0.10$), each of which tend to bias estimates based
on strong emission lines toward high oxygen abundances.
These spectra all fall in a regime where the ``standard'' strong line
methods for metallicity determinations are not well calibrated either
empirically or by photoionization modeling.
By comparing our spectra directly to photoionization models, we estimate
oxygen abundances in the range of $7.9\leq12 + \log{\mbox{(O/H)}}\leq8.4$,
consistent with the scatter of the mass-metallicity relation.
We discuss the physical nature of these galaxies that leads to their unusual
spectra (and previous classification as outliers), finding their low excitation,
elevated N/O, and strong Balmer absorption are consistent with the properties
expected from galaxies evolving past the ``Wolf-Rayet galaxy'' phase.
We compare our results to the ``main" sample of \cite{peeples08} and
conclude that they are outliers primarily due to enrichment of nitrogen relative 
to oxygen, and not due to unusually high oxygen abundances for their masses 
or luminosities.
\end{abstract}



\section{INTRODUCTION}\label{sec:intro}

There is a fundamental relationship between the mass of stars in a galaxy 
and its metallicity evolution (hereafter, the M-Z relation). 
Empirically, this has been observed as a luminosity-metallicity 
relationship for low redshift dwarf galaxies
\citep[e.g.,][and references therein]{lequeux79, skillman89, lee06b} 
and spiral galaxies 
\citep[e.g.,][and references therein]{mccall85, garnett87, zaritsky94, tremonti04}. 
This robust relationship is observed over a range of 10 magnitudes in 
galaxy optical luminosity \citep[e.g.,][]{zaritsky94, tremonti04, lee06b}. 
In recent years, galaxies at higher redshifts have also shown a 
mass-metallicity or luminosity-metallicity relationship, and mounting 
evidence suggests this relationship evolves with time 
\citep[e.g.,][and references therein]{kobulnicky03, kobulnicky04, 
shapley04, savaglio05, maiolino08}, but see also \citet{Mann10}. 
Thus, this relationship provides both a very strong constraint 
on theories of galaxy evolution and a tool to better understand 
galaxies at higher redshifts.

\citet[hereafter T04]{tremonti04} derived luminosities, metallicities, 
and masses for $\sim$53,000 low redshift galaxies observed in the 
Sloan Digital Sky Survey \citep[SDSS;][]{york00} and convincingly 
demonstrated that the basis of the empirically observed luminosity-metallicity 
relationship is an underlying association between stellar mass and metal abundance. 
The metallicities were estimated by fitting all observable strong emission 
lines and deriving a metallicity likelihood distribution for each galaxy 
based on theoretical model fits calculated using a combination of stellar 
population synthesis and photoionization models. 
These models were then used with z-band luminosities to form mass estimates. 

The physical driver for the M-Z relationship is still debated.  
Many studies favor supernova driven winds for the inability of a 
low-mass galaxy to retain its newly synthesized heavy elements, resulting 
in a lower effective yield with decreasing mass \citep[e.g.,][]{dekel86}. 
Other observational and theoretical studies are critical of this theory.
For example, \citet{dalcanton07} emphasizes the additional importance 
of star formation efficiency as outflows are an insufficient 
regulator in the absence of depressed star formation. 
Thus, a better understanding of the mass-metallicity relationship remains important. 

Although the M-Z relationship is well defined, it shows measurable scatter.  
Since observational error accounts for only half of the 
metallicity spread in the M-Z relation \citep{cooper08}, one or 
more physical processes may be responsible for the remainder.  
Suggestions for the scatter include variations in the star formation 
history \citep[e.g., recent starbursts,][]{contini02}, variations in 
stellar surface mass density \citep{ellison08}, and variations in local 
galaxy density \citep[e.g.,][and references therein]{cooper08}. 
Motivated by the fact that outliers often provide key insights into the nature of 
physical relationships \citep[e.g.,][]{skillman96}, \citet[hereafter P08]{peeples08} 
and \citet{peeples09} identified samples of outlier galaxies from the T04 study. 

P08 analyzed a sample of 41 high oxygen abundance, low-luminosity 
galaxy outliers from the M-Z relation of T04. 
Their ``main" sample is comprised of 24 high abundance 
(8.95 $\le$ 12 + log(O/H) $\le$ 9.27), relatively low-mass 
(9.1 $\le$ log$(\mbox{M}_{\star}/\mbox{M}_{\odot})$ $\le$ 9.9), 
low-luminosity (sub-L$_\star$; $-$19 $\le$ M$_{B}$ $\le$ $-17$) dwarf galaxies. 
A redshift lower limit of z $>0.024$ was imposed to ensure the 
inclusion of the [\ion{O}{2}] $\lambda\lambda3726,3729$ emission 
doublet in the SDSS spectral coverage\footnote{Seven galaxies with z $< 0.024$ 
passed the subsequent error and visual inspection cuts imposed by \cite{peeples08} 
and so were kept in their ``main" sample.}. 
The data set was extended to include even lower mass galaxies, creating 
a second ``very low mass" sample, with 
7.4 $\le$ log$(\mbox{M}_{\star}/\mbox{M}_{\odot})$ $\le$ 9.0, 
$-$17 $\le$ M$_{B}$ $\le$ $-14$, and 8.68 $\le$ 12 + log(O/H) $\le$ 9.12. 
In order to increase the sample to these lower masses, the redshift limit was dropped. 
All galaxies in the resulting ``very low mass" sample have z $<0.024$, and 
thus lack an [\ion{O}{2}] $\lambda\lambda3726,3729$ spectral measurement. 
The typical oxygen abundance offset in the O/H - M$_B$ plane for a galaxy in 
the ``main'' sample is $\approx$ 0.4 dex, while the typical offset for the ``very 
low mass" sample is $\approx$ 0.6 dex (with offsets as large as $\approx$ 0.9 dex).

P08 favored isolated or undisturbed systems with relatively low gas mass 
fractions nearing the end of their star formation activity as an explanation 
for the unexpectedly high oxygen abundances.
Other possible causes for the high abundance outliers were considered and 
ruled out, such as discrepantly low luminosities for their masses or 
inaccurate metallicity calculations. 

Nebular oxygen abundances derived from the observations of strong
emission lines and in the absence of direct measurements of 
the electron temperature are always subject to systematic effects
\citep[e.g.,][and references therein]{kbg03}.  This is particularly
true at the higher abundances typically found in spiral galaxies
where several calibrations of methods based solely on strong lines
result in systematically higher oxygen abundances when compared to
oxygen abundances derived from direct measurements of the electron 
temperature \citep[see, e.g., discussion in][and references therein]{bresolin07}.
Importantly, \cite{yin07} and \cite{perez-montero09} have pointed out that 
strong line calibrators that are based on the strength of the [\ion{N}{2}]
$\lambda\lambda$6548,6584 emission lines are biased in the sense that large
values of N/O lead to overestimates of the oxygen abundance.
At lower values of oxygen abundance, in general, 
the strong line methods show better agreement 
with the oxygen abundances derived from direct temperature measurements,
but \cite{vanzee06b} have shown 
that there can be significant discrepancies at low values of 
excitation ($\lambda$5007/$\lambda$3727).  Given the uncertainties
in strong line oxygen abundance measurements, it is warranted to 
reinvestigate the conclusions of P08.

The prospect of galaxy outliers from the M-Z relationship, and their consequences 
for galaxy evolution models, motivated the re-analysis of these objects.
In this paper we discuss the previous SDSS and new MMT observations in 
\S~\ref{sec:data} and describe the analysis of the latter in \S~\ref{sec:analysis}.
In \S~\ref{sec:SDSS} we use our new observations of the
[\ion{O}{2}] $\lambda\lambda3726,3729$ emission lines to compare the properties
of the outliers to the SDSS galaxies.
Section~\ref{sec:metallicity} is dedicated to looking at several metallicity 
determinations and the appropriate applications, including the O3N2 method (\S~\ref{sec:O3N2}), 
the N2 indicator (\S~\ref{sec:N2}), and the R$_{23}$ index (\S~\ref{sec:R23}).
Our best estimates of the oxygen abundances, the nature of the objects in the 
``very low mass" sample that gives rise to their discrepant spectra, and an inspection 
of the nature of the objects in the P08 ``main" sample are discussed in 
\S~\ref{sec:discussion}.


\section{DATA}\label{sec:data}

\subsection{Sample}\label{sec:obs}
Four of the 17 metal-rich galaxies identified by P08 as ``very low mass" 
($\log(\mbox{M}_{\star}$/M$_{\odot})<8.7$) outliers from the M-Z relationship 
were selected for follow-up observations with the MMT (see Table~\ref{tbl:1}).  
These targets were chosen both for their significant departures from the mass-metallicity 
relationship and their availability during a single scheduled observing run. 
They have suggested high metallicities, as measured by T04, 
of $8.69 \leq 12+\log(\mbox{O/H}) \leq 8.86$. 
For SDSS J022628.28+010937.7, SDSS J024121.80+000329.2, SDSS 
J082639.19+253553.5, and SDSS J082633.77+252959.2 respectively
(hereafter abbreviated as SDSS- plus the first six digits of the Right Ascension), 
these oxygen abundances lie 0.37, 0.40, 0.55, and 0.55 dex above 
(and well outside the $\sim$ 0.1 dex scatter of) the luminosity-metallicity 
relationship histogram medians presented by P08. 
Like all of the objects in the P08 parent sample, these four galaxies are fairly 
isolated\footnote{Note that J082639.19+253553.5 and J082633.77+252959.2 
are members of the same group (Peeples, private communication, 2010).}, 
low-redshift ($0.0051<z<0.0227$) dwarfs ($-17<\mbox{M}_{B}<-15$), 
with no obvious companions and somewhat depressed star formation rates (P08).


\begin{deluxetable}{cccccccc}
\tablewidth{0pt}
\tabletypesize{\footnotesize}
\tablecaption{Present Sub-Sample of Low-Luminosity Outliers from Peeples et al.\ 2008}
\tablehead{
\colhead{Object} & \colhead{SDSS Spectra} & \colhead{RA} & \colhead{DEC} & \colhead{M$_{B}$} & \colhead{log M$_{\star}$} & \colhead{Redshift} & \colhead{g-r}}
\startdata
{SDSS J022628.28+010937.7} & {spSpec-51869-0406-561.fits} & {36.6179}   & {1.16053}     & {-16.86} & {7.92} & {0.0051} & {0.51} \\
{SDSS J024121.80+000329.2} & {spSpec-52177-0707-355.fits} & {40.3408}   & {0.05813}     & {-16.56} & {8.70} & {0.0227} & {0.46} \\
{SDSS J082639.19+253553.5} & {spSpec-52945-1586-164.fits} & {126.6633} & {25.59821}  & {-15.87} & {8.59} & {0.0078} & {0.71} \\
{SDSS J082633.77+252959.2} & {spSpec-52945-1586-161.fits} & {126.6407} & {25.49979}  & {-15.23} & {8.07}  & {0.0072} & {0.55} \\
\enddata
\tablecomments{Publicly available SDSS spectra can be found at SDSS.org. 
The RA, DEC, M$_{B}$, log M$\star$, and redshift values are taken from P08 
(see their Table 1 for more details). The SDSS DR 7 provided g-r colors. }
\label{tbl:1}
\end{deluxetable}


\begin{figure}
\epsscale{1.0}
\plotone{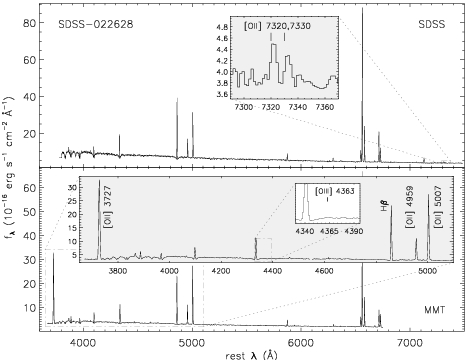}
\caption{Spectra of SDSS J022628.28+010937.7. 
The two panels compare the optical \ion{H}{2} region spectra from the SDSS and the MMT. 
Note that the SDSS spectral bandpass does not contain [OII] $\lambda$3727, 
but does encompass measurable [\ion{O}{2}] $\lambda\lambda$7320,7330, 
as shown in the upper inset box. 
The bottom inset box is an enlarged view of the MMT spectrum to emphasize the 
dominance of [\ion{O}{2}] over [\ion{O}{3}]. 
This scale also reveals the significant [\ion{N}{2}] $\lambda\lambda$6548,6584 strength 
and the lack of a [\ion{O}{3}] $\lambda$4363 detection.}
\label{fig:1}
\end{figure}


\begin{figure}
\epsscale{1.0}
\plotone{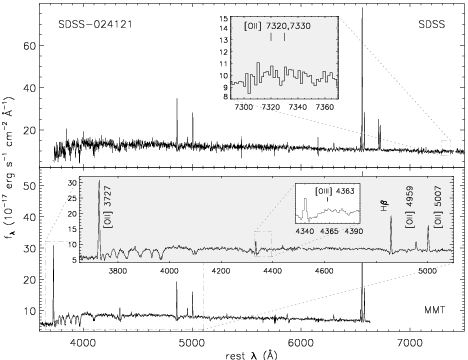}
\caption{Same as Figure~\ref{fig:1}, but the spectra are of SDSS J024121.80+000329.2. 
The SDSS spectrum is too noisy to detect the [\ion{O}{2}] $\lambda\lambda$7320,7330 doublet,
but displays the strong nitrogen lines. 
Again, note the strength of the [\ion{O}{2}] $\lambda$3727 emission relative to the [\ion{O}{3}] 
emission in the MMT spectrum.}
\label{fig:2}
\end{figure}


\begin{figure}
\epsscale{1.0}
\plotone{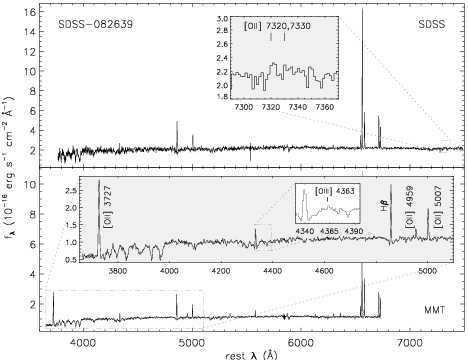}
\caption{Same as Figure~\ref{fig:1}, but the spectra are of SDSS J082639.19+253553.5. 
The SDSS spectrum falls short of capturing the [\ion{O}{2}] $\lambda$3727 line 
and lacks the high signal to noise needed to detect the red [\ion{O}{2}] lines. 
Below, in the MMT spectrum, the [\ion{O}{2}] $\lambda$3727 line strength exceeds that 
of the [\ion{O}{3}] $\lambda\lambda$4959,5007 doublet.
Also notice the strong [\ion{N}{2}] $\lambda\lambda$6548,6584 lines.}
\label{fig:3}
\end{figure}


\begin{figure}
\epsscale{1.0}
\plotone{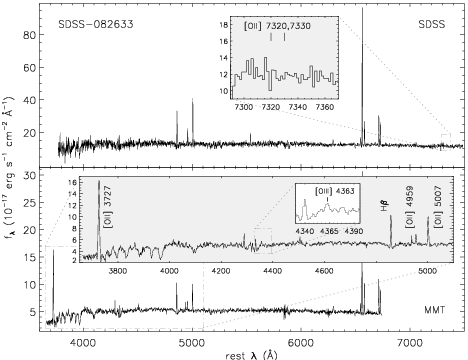}
\caption{Same as Figure~\ref{fig:1}, but the spectra are of SDSS J082633.77+252959.2. 
The MMT spectrum for this object has the most prominent [\ion{O}{2}] $\lambda$3727 
line relative to the [\ion{O}{3}] doublet and, again, strong [\ion{N}{2}] lines.
Note that while [\ion{O}{3}] $\lambda$4363 appears to be measurable, 
its flux is less than the 3 $\sigma$ uncertainty, and so was not used.}
\label{fig:4}
\end{figure}


\subsection{SDSS Spectra}\label{sec:sdss_spectra}

The measurements made by T04 and used by P08 were derived from the publicly 
available SDSS\footnote{http://www.sdss.org/dr4/} 
Data Release 4 \citep{SDSSDR4} data files referenced in Table~\ref{tbl:1}.
We used the SDSS pipeline reduced spectra (rather than performing our own 1-D 
extraction and reductions) to minimize differences between analyses. 
For a thorough description of the data reduction refer to \citet{SDSSDR4}.
While the median signal-to-noise (S$/$N) values for the SDSS spectra (12--47) 
meet the S/N (per pixel) $> 8$ requirement for reliable metallicity estimates 
\citep{kobulnicky99}, the wavelength coverage (3800--9200 \AA) excludes the 
[\ion{O}{2}] $\lambda$3727 emission line from their spectra.
Although the red [\ion{O}{2}] $\lambda\lambda$7320,7330 lines are included 
in the wavelength range, they are only detected in the spectrum of SDSS-022628.
In Figures~\ref{fig:1}-\ref{fig:4}, we have indicated the locations of the [\ion{O}{2}] 
$\lambda$3727 and [\ion{O}{2}] $\lambda\lambda$7320,7330 emission lines. 

In \S ~\ref{sec:SDSS} and later in the paper, we compare galaxies from P08 to
samples from the SDSS.  In these comparisons,
we use values of nebular emission line strengths, stellar absorption line strengths,
star formation rates, and
stellar masses based on SDSS spectra and photometry from the MPA-JHU data
catalogue\footnote{http://www.mpa-garching.mpg.de/SDSS/}.
Stellar masses were determined based on fits to photometry following the work of \citet{kauffmann03},
and star formation rates (SFR) were based on \citet{brinchmann04} and
\citet{salim07}\footnote{MPA-JHU used a method similar to \cite{salim07} to aperture correct their SFRs.}.
Objects with [\ion{O}{2}], [\ion{O}{3}], H$\beta$, [\ion{N}{2}], and H$\alpha$ line strengths
less than 5 $\sigma$ were filtered out in order to increase the quality of this data set.


\subsection{MMT Spectra}\label{sec:mmt}
\subsubsection{Observations}\label{sec:mmtobs}
New MMT observations were acquired in order to obtain improved S$/$N spectra and 
extended blue wavelength coverage including the [\ion{O}{2}] $\lambda$3727 line. 
The MMT data were taken with the Blue Channel spectrograph 
\citep{schmidt89} on the UT date of 2008 November 1-2. 
Sky conditions were optimal with no cloud cover and sub-arcsecond seeing. 
A 500 line grating, $1\arcsec$ slit, and UV-36 blocking filter were used, 
yielding an approximate dispersion of 1.2 \AA\ per pixel, a full width at half 
maximum resolution of $\lesssim3$ \AA, and a wavelength coverage of 3690--6790 \AA. 
Bias frames, flat-field lamp images, and sky flats were taken each night. 
The latter were primarily necessary due to significant differences between 
the chip illumination patterns of the sky and the MMT Top Box that houses 
the Blue Channel incandescent flat-field lamp. 
Multiple standard stars from \citet{oke90} with spectral energy distributions 
peaking in the blue and containing minimal absorption were observed throughout 
the night using a 5$\arcsec$ slit over a range of airmasses. 

All four galaxies had strong central brightness peaks which were centered 
on the $1\arcsec\times180\arcsec$ slit. 
Three 900 second exposures (600 seconds for SDSS-086239) were made at a 
fixed position angle which approximated the parallactic angle at half 
the total integration time. 
This, in addition to observing the galaxies at an airmass below 1.5, 
served to minimize the wavelength-dependent light loss due to 
differential refraction \citep{filippenko82}. 
A single slit position for each target was sufficient to characterize 
the global oxygen abundance, as metallicity gradients are small in 
low-mass galaxies \citep[e.g.,][]{skillman89, ks96, ks97, lee06a}. 
Finally, combined helium, argon, and neon arc lamps were observed at 
each pointing for accurate wavelength calibration. 


\subsubsection{Data Reduction}\label{sec:mmtreduct}
The MMT observations were processed using ISPEC2D \citep{moustakas06}, 
a long-slit spectroscopy data reduction package written in IDL. 
A master bias frame was created from $\gtrsim 20$ zero second exposures 
by discarding the highest and lowest value at each pixel and taking the median. 
Master sky and dome flats were similarly constructed after normalizing 
the counts in the individual images. 
Those calibration files were then used to bias-subtract, flat-field, 
and illumination-correct the raw data frames. 
Dark current was measured to be an insignificant $\sim 1$ e$^{-}$ 
per pixel per hour and was not corrected for. 

Misalignment between the trace of the light in the dispersion direction and the 
orientation of the CCD detector was rectified via the mean trace of the standard 
stars for each night, providing alignment to within a pixel across the detector. 
A two-dimensional sky subtraction was performed using individually selected 
sky apertures, followed by a wavelength calibration applied from the HeArNe 
comparison lamps taken at the same telescope pointing. 
Airmass dependent atmospheric extinction and reddening were corrected for 
using the standard Kitt Peak extinction curve \citep{crawford70}. 

For each galaxy, the three sub-exposures were combined, 
eliminating cosmic rays in the process. 
The resulting images were then flux-calibrated using the sensitivity curve 
derived from the standard star observations taken throughout a given night. 
Finally, the trace fit to the strongest continuum source in the slit was used 
to extract the galaxy light within 15-30$\arcsec$ apertures that encompassed 
$\gtrsim$ 99\% of the spatially smooth and centrally peaked emission.
Figures~\ref{fig:1}-\ref{fig:4} show the resulting one-dimensional spectra 
(with median S/N values $>$ 30) in comparison to the SDSS spectra. 
Inset windows with a narrower spectral range emphasize the blue emission lines 
and the dominance of [\ion{O}{2}] over [\ion{O}{3}] in all four galaxies.


\section{ANALYSIS}\label{sec:analysis}


\subsection{Emission Line Measurements}\label{sec:iraf}
Emission line strengths were measured for both the MMT and SDSS spectra 
using standard methods available within 
IRAF\footnote{IRAF is distributed by the National Optical Astronomy 
Observatories, which are operated by the Association of Universities for Research 
in Astronomy, Inc., under cooperative agreement with the National Science Foundation.}. 
In particular, the SPLOT routine was used to analyze the extracted one-dimensional 
spectra and fit Gaussian profiles to emission lines to determine their integrated fluxes. 
The H$\alpha$ and the adjacent [\ion{N}{2}] lines were fitted 
simultaneously for an accurate deblending. 

Special attention was paid to the Balmer lines, which are located in 
troughs of significant underlying stellar absorption. 
Although the equivalent widths of the H$\alpha$ emission lines were large 
enough that the underlying absorption was not a concern, 
this was not the case for H$\beta$ and the bluer Balmer lines. 
We experimented with several different methods to correct the measurement of the 
H$\beta$ emission flux for the effects of underlying absorption. 
At one extreme, maximizing the H$\beta$ line intensity 
(by overestimating the underlying absorption) 
leads to a minimum H$\alpha$/H$\beta$ value and a minimum estimate of the reddening. 
Because the underlying absorption is much broader than the emission and these 
features are both well resolved and relatively high signal-to-noise, the subtraction 
of the underlying absorption represents only a small contribution to the uncertainty 
in the reddening corrections (see Table~\ref{tbl:2}).

\begin{figure}
   \begin{center}
      \begin{tabular}{cc}
          \resizebox{85mm}{!}{\includegraphics{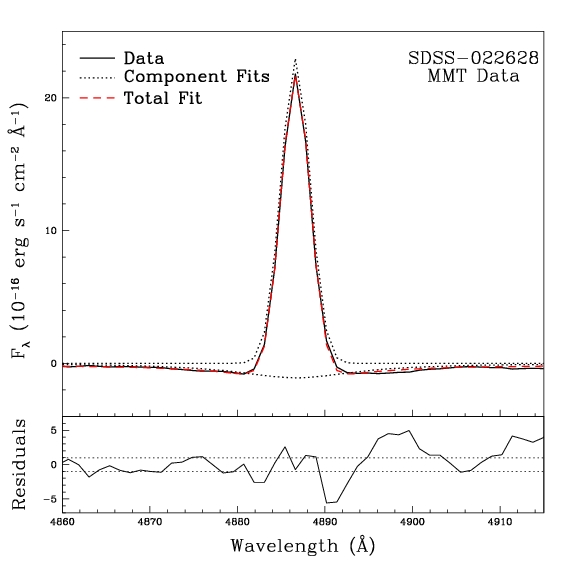}} & \resizebox{85mm}{!}{\includegraphics{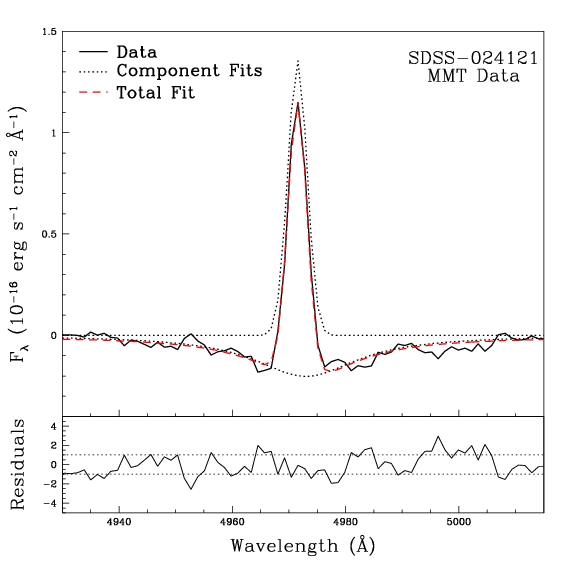}} \\
          \resizebox{85mm}{!}{\includegraphics{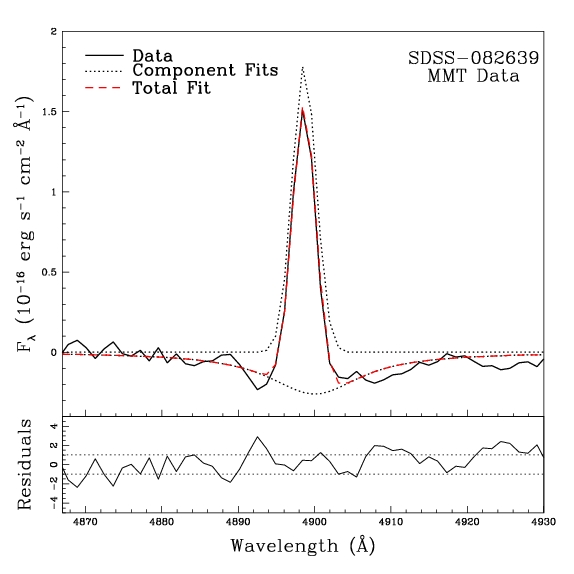}} & \resizebox{85mm}{!}{\includegraphics{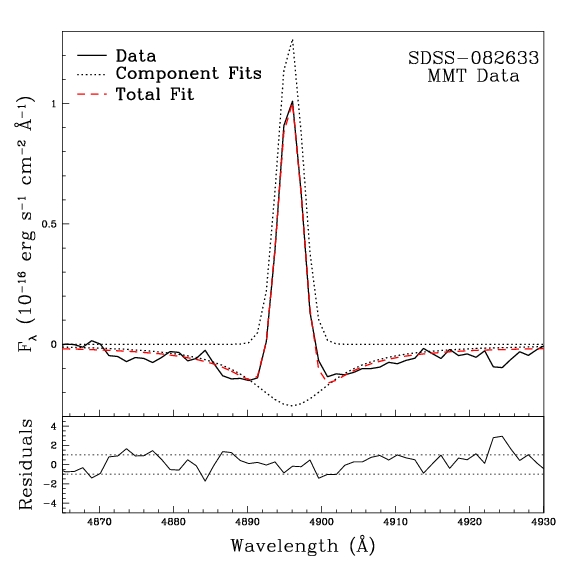}} \\
       \end{tabular}    
       \caption{H$\beta$ region of the MMT spectra, as measured in PAN. 
       The dotted gray lines correspond to the Lorentzian absorption and Gaussian emission component fits. 
       The red dashed line displays the best overall fit to the original data (solid black line). 
       The residual difference between the fit and data is shown in the bottom panel, with very little 
       divergence near the emission peak, confirming a successful fit. 
       The ensuing H$\beta$ emission flux, accounting for Balmer absorption, can be found in Table~\ref{tbl:2}.}
     \label{fig:5}
   \end{center}
\end{figure}

For our final analysis of the Balmer emission lines, we used the PAN\footnote{PAN 
was written by Rob Dimeo as a part of the Data Analysis and Visualization Environment, 
which is a software package developed at the NIST Center for Neutron Research for 
the reduction, visualization, and analysis of inelastic neutron scattering data, and was 
funded by the National Science Foundation.} analysis package to simultaneously fit the continuum, 
the Gaussian emission peak, and a broad, negative Lorentzian absorption feature. 
PAN uses a least-squares fit to minimize $\chi^{2}$ and estimates the uncertainty using 
a ``bootstrap" Monte Carlo error analysis, providing a reliable measurement of the flux 
and associated uncertainty of the Balmer emission line. 
Figure~\ref{fig:5} exhibits both the individual component and total fits determined by PAN, 
with minimal residuals validating the goodness of fit to our spectra. 
The emission line fluxes are reported relative to H$\beta$ in Table~\ref{tbl:2},
and represent the multiple component fits for the H$\beta$ and H$\gamma$ emission lines and 
the single or deblended Gaussian profile fits from SPLOT for the rest of the emission lines.


The errors of the flux measurements were approximated using 
\begin{equation}
	\sigma \approx 2 \sqrt{N}\ rms ,	\label{eq:uncertainty}
\end{equation}
where N is the number of pixels spanning the Gaussian profile 
fit to the narrow emission lines (typically 11). 
The rms noise in the continuum was taken to be the average of 
the rms on each side of an emission line. 
This error approximation is valid for weak lines whose uncertainty 
is dominated by error from the continuum subtraction. 
For cases where the flux measurements were much stronger than 
the rms noise of the continuum, the error is dominated by 
flux calibration and de-reddening uncertainty. 
This is true for several lines in both the SDSS and MMT spectra of SDSS-022628, 
and for all the H$\alpha$ lines (where a minimum uncertainty of 2\% was assumed). 
Flux line strengths relative to H$\beta$ and corresponding 
errors are listed in Table \ref{tbl:2}. 
As we were not able to detect [\ion{O}{3}] $\lambda$4363 and \ion{He}{2} 
$\lambda$4686 at the level of more than 3 $\sigma$ in any of the spectra, 
a flux upper limit was estimated using Equation~\ref{eq:uncertainty}.  
We calculated upper limits on the electron temperatures based 
on these $\lambda$4363 fluxes, but none provided significant 
constraints for abundance calculations.
In some cases, the H$\alpha$/H$\beta$ ratios for the SDSS 
and MMT spectra are significantly different.
Using the MMT data for our reference ratio, the average percentage
difference between the SDSS and MMT H$\alpha$/H$\beta$ ratio 
is 15\%, whereas the average 
[\ion{O}{3}] $\lambda$5007/H$\beta$ percentage difference is 16\%.
Since the [\ion{O}{3}]/H$\beta$ ratio is less sensitive to flux calibrations,
these differences between the two data sets are likely due to 
using long-slit (MMT spectra) versus circular fiber apertures (SDSS spectra).


\subsection{Reddening Corrections}\label{sec:redcor}
The wide range of observed wavelengths require fluxes to 
be corrected for extinction and reddening. 
Since the relative intensities of the Balmer lines are nearly independent 
of both density and temperature, they can be used to solve for the reddening. 
Assuming standard \ion{H}{2} region characteristics (T$_{e}=1.25 \times 10^{4}$ K 
and n$_{e}=10^{2}$ cm$^{-3}$), both the MMT and SDSS spectra were de-reddened 
using a Balmer decrement of 2.82 \citep{storey87} and the \citet{cardelli89} 
reddening law (with A$_{V}=3.1\ E(B-V)$). 
The extinction, $A_{1}(\lambda)$, from the \citet{cardelli89} law was 
calculated using the York Extinction Solver 
\citep{mccall04}\footnote{http://www1.cadc-ccda.hia-iha.nrc-cnrc.gc.ca/community/YorkExtinctionSolver/}. 
With these values we then derived the reddening value, $E(B-V)$, using
\begin{equation}
	\log{\frac{I(H\alpha)}{I(H\beta)}\ } = \log{\frac{F(H\alpha)}{F(H\beta)}\ } + 0.4\ E(B-V)\ [A_{1}(H\alpha)-A_{1}(H\beta)],
	\label{eq:dered} 
\end{equation}
where F(H$\alpha$)/F(H$\beta$) is the observed flux ratio and 
I(H$\alpha$)/I(H$\beta$) is the de-reddened line intensity ratio using 
case B from \citet{storey87}. 
Following \citet{lee04} the reddening value can be converted 
to the logarithmic extinction at $H\beta$ as
\begin{equation}
	c(H\beta) = 1.43\ E(B-V).
	\label{eq:cHbeta}
\end{equation}
Original and de-reddened flux values for both the MMT and SDSS data are given 
in Table~\ref{tbl:2}, where errors were propagated from those associated 
with the individual line measurements. 
All of the H$\alpha$/H$\beta$ ratios are larger than 2.82, 
indicative of significant extinction and reddening due to dust. 
Since the Galactic latitudes for the four galaxies are all large 
(ranging from $31 \degr$  to $54 \degr$), 
little foreground extinction from Galactic dust is expected, and, indeed, 
the calculated values of $E(B-V)$ (see Table~\ref{tbl:2}) are substantially 
greater than the foreground extinction determined by \citet{schlegel98}.
Our reddening corrections can be checked by comparing the corrected 
H$\gamma$/H$\beta$ ratios with their theoretical values. 
As seen in Table~\ref{tbl:2}, this ratio is consistent with the theoretical 
case B recombination ratio of 0.47 \citep{storey87} in all four cases. 
This result implies accurate reddening corrections were made,
which strengthens the argument for significant intrinsic extinction. 


\begin{deluxetable}{cccccccccccc}
\tabletypesize{\scriptsize}
\rotate
\tablecaption{Normalized and Reddening Corrected Emission Line Intensity Ratios for the Observed Sample\label{tbl:2} }
\tablewidth{0pt}
\tablehead{
\colhead{} & \multicolumn{2}{c}{SDSS-022628} && \multicolumn{2}{c}{SDSS-024121} && \multicolumn{2}{c}{SDSS-082639} && \multicolumn{2}{c}{SDSS-082633} \\
\colhead{MMT Data} & \colhead{F($\lambda$)/F(H$\beta$)} & \colhead{I($\lambda$)/I(H$\beta$)} && \colhead{F($\lambda$)/F(H$\beta$)} & \colhead{I($\lambda$)/I(H$\beta$)} &&
\colhead{F($\lambda$)/F(H$\beta$)} & \colhead{I($\lambda$)/I(H$\beta$)} && \colhead{F($\lambda$)/F(H$\beta$)} & \colhead{I($\lambda$)/I(H$\beta$)}}
\startdata
{[O~II] $\lambda$3727}	& 2.12$\pm$0.06 	& 2.47$\pm$0.09		&& 2.62$\pm$0.13	 & 4.13$\pm$0.22	 && 1.89$\pm$0.12	  & 3.67$\pm$0.25	  && 3.50$\pm$0.17	   & 3.90$\pm$0.22	   \\
H$\gamma$ $\lambda$4340	& 0.40$\pm$0.01 	& 0.43$\pm$0.01		&& 0.33$\pm$0.04	 & 0.41$\pm$0.05	 && 0.37$\pm$0.06	  & 0.49$\pm$0.08	  && 0.34$\pm$0.05	   & 0.38$\pm$0.05	   \\
{[O~III] $\lambda$4363}	& $<$ 0.02 (3 $\sigma$)	& $<$ 0.02 (3 $\sigma$)	&& $<$ 0.12 (3 $\sigma$) & $<$ 0.14 (3 $\sigma$) && $<$ 0.15 (3 $\sigma$) & $<$ 0.19 (3 $\sigma$) && $<$ 0.16 (3 $\sigma$) & $<$ 0.17 (3 $\sigma$) \\ 
{He~II $\lambda$4686}	& $<$ 0.01 (3 $\sigma$)	& $<$ 0.01 (3 $\sigma$)	&& $<$ 0.10 (3 $\sigma$) & $<$ 0.10 (3 $\sigma$) && $<$ 0.15 (3 $\sigma$) & $<$ 0.16 (3 $\sigma$) && $<$ 0.10 (3 $\sigma$) & $<$ 0.10 (3 $\sigma$) \\
H$\beta$ $\lambda$4861	& 1.00$\pm$0.03		& 1.00$\pm$0.03		&& 1.00$\pm$0.04	 & 1.00$\pm$0.04	 && 1.00$\pm$0.03	  & 1.00$\pm$0.03	  && 1.00$\pm$0.06	   & 1.00$\pm$0.06	   \\
{[O~III] $\lambda$4959}	& 0.39$\pm$0.01		& 0.39$\pm$0.01		&& 0.28$\pm$0.04	 & 0.27$\pm$0.03	 && 0.18$\pm$0.05	  & 0.17$\pm$0.05	  && 0.32$\pm$0.08	   & 0.31$\pm$0.08	   \\
{[O~III] $\lambda$5007}	& 1.18$\pm$0.03		& 1.16$\pm$0.03		&& 0.70$\pm$0.04	 & 0.67$\pm$0.04	 && 0.54$\pm$0.05	  & 0.50$\pm$0.05	  && 0.95$\pm$0.09	   & 0.93$\pm$0.09	   \\
{He~I $\lambda$5876}	& 0.12$\pm$0.02		& 0.11$\pm$0.02		&& 0.15$\pm$0.02	 & 0.11$\pm$0.02	 && 0.23$\pm$0.04	  & 0.15$\pm$0.03	  && 0.13$\pm$0.02	   & 0.13$\pm$0.02	   \\
{[N~II] $\lambda$6548}	& 0.20$\pm$0.01		& 0.17$\pm$0.01		&& 0.35$\pm$0.04	 & 0.23$\pm$0.02	 && 0.42$\pm$0.05	  & 0.23$\pm$0.03	  && 0.19$\pm$0.05	   & 0.15$\pm$0.04	   \\
H$\alpha$ $\lambda$6563	& 3.24$\pm$0.10		& 2.79$\pm$0.08		&& 4.30$\pm$0.15	 & 2.79$\pm$0.10	 && 5.19$\pm$0.15	  & 2.79$\pm$0.10	  && 3.50$\pm$0.16	   & 2.79$\pm$0.13	   \\
{[N~II] $\lambda$6584}	& 0.63$\pm$0.02		& 0.55$\pm$0.02		&& 1.02$\pm$0.05	 & 0.67$\pm$0.03	 && 1.38$\pm$0.06	  & 0.74$\pm$0.04	  && 0.71$\pm$0.06	   & 0.57$\pm$0.04	   \\
\hline
{H$\beta$ Flux}\tablenotemark{1} & \multicolumn{2}{c}{956$\pm$19}   && \multicolumn{2}{c}{57.7$\pm$1.7}	 && \multicolumn{2}{c}{70.3$\pm$1.4}  && \multicolumn{2}{c}{55.4$\pm$2.3}  \\
{H$\beta$ EW}\tablenotemark{2} 	& \multicolumn{2}{c}{-23.6$\pm$0.7} && \multicolumn{2}{c}{-6.1$\pm$0.8}	 && \multicolumn{2}{c}{-4.9$\pm$0.9}  && \multicolumn{2}{c}{-5.1$\pm$0.9}  \\
{H$\beta$ Abs EW}		& \multicolumn{2}{c}{5.7$\pm$0.2}   && \multicolumn{2}{c}{8.1$\pm$0.5}	 && \multicolumn{2}{c}{6.9$\pm$0.3}   && \multicolumn{2}{c}{7.0$\pm$0.4}   \\
{H$\alpha$ EW}			& \multicolumn{2}{c}{-87.2$\pm$1.8} && \multicolumn{2}{c}{-34.5$\pm$2.5} && \multicolumn{2}{c}{-25.7$\pm$1.4} && \multicolumn{2}{c}{-24.9$\pm$1.8} \\
{E(B-V)}                   	& \multicolumn{2}{c}{0.14$\pm$0.01} && \multicolumn{2}{c}{0.42$\pm$0.02} && \multicolumn{2}{c}{0.62$\pm$0.03} && \multicolumn{2}{c}{0.22$\pm$0.01} \\
{C(H$\beta$)}            	& \multicolumn{2}{c}{0.20$\pm$0.01} && \multicolumn{2}{c}{0.61$\pm$0.02} && \multicolumn{2}{c}{0.88$\pm$0.03} && \multicolumn{2}{c}{0.31$\pm$0.01} \\

\hline \\
{} & \multicolumn{2}{c}{SDSS-022628} && \multicolumn{2}{c}{SDSS-024121} && \multicolumn{2}{c}{SDSS-082639} && \multicolumn{2}{c}{SDSS-082633} \\
{SDSS Data} & {F($\lambda$)/F(H$\beta$)} & {I($\lambda$)/I(H$\beta$)} && {F($\lambda$)/F(H$\beta$)} & {I($\lambda$)/I(H$\beta$)} &&
{F($\lambda$)/F(H$\beta$)} & {I($\lambda$)/I(H$\beta$)} && {F($\lambda$)/F(H$\beta$)} & {I($\lambda$)/I(H$\beta$)} \\
\hline      
H$\beta$ $\lambda$4861	& 1.00$\pm$0.03	& 1.00$\pm$0.03	&& 1.00$\pm$0.11 & 1.00$\pm$0.11 && 1.00$\pm$0.09 & 1.00$\pm$0.09 && 1.00$\pm$0.09 & 1.00$\pm$0.09 \\
{[O~III] $\lambda$4959}	& 0.39$\pm$0.02	& 0.38$\pm$0.02	&& 0.18$\pm$0.07 & 0.17$\pm$0.07 && \nodata	  & \nodata	  && 0.37$\pm$0.07 & 0.36$\pm$0.07 \\
{[O~III] $\lambda$5007}	& 0.92$\pm$0.03	& 0.88$\pm$0.03	&& 0.61$\pm$0.09 & 0.59$\pm$0.08 && 0.42$\pm$0.08 & 0.39$\pm$0.07 && 1.00$\pm$0.09 & 0.96$\pm$0.09 \\
{[N~II] $\lambda$6548}	& 0.28$\pm$0.01	& 0.19$\pm$0.01	&& 0.30$\pm$0.04 & 0.23$\pm$0.03 && 0.42$\pm$0.05 & 0.22$\pm$0.02 && 0.22$\pm$0.05 & 0.15$\pm$0.03 \\
H$\alpha$ $\lambda$6563	& 4.09$\pm$0.13	& 2.79$\pm$0.09	&& 3.71$\pm$0.29 & 2.79$\pm$0.22 && 5.34$\pm$0.34 & 2.79$\pm$0.18 && 4.02$\pm$0.28 & 2.79$\pm$0.19 \\
{[N~II] $\lambda$6584}	& 0.87$\pm$0.03	& 0.60$\pm$0.02	&& 0.95$\pm$0.08 & 0.72$\pm$0.06 && 1.46$\pm$0.10 & 0.77$\pm$0.05 && 0.74$\pm$0.07 & 0.52$\pm$0.05 \\
{[O~II] $\lambda$7320}	& 0.03$\pm$0.01	& 0.02$\pm$0.01	&& \nodata	 & \nodata	 && \nodata	  & \nodata	  && \nodata	   & \nodata	   \\
{[O~II] $\lambda$7330}	& 0.02$\pm$0.01	& 0.01$\pm$0.01	&& \nodata	 & \nodata	 && \nodata	  & \nodata	  && \nodata	   & \nodata	   \\
\hline
{H$\beta$ Flux} 	& \multicolumn{2}{c}{979$\pm$24}    && \multicolumn{2}{c}{87.2$\pm$6.6}  && \multicolumn{2}{c}{114.7$\pm$7.2} && \multicolumn{2}{c}{88.0$\pm$5.8}  \\
{H$\beta$ EW}		& \multicolumn{2}{c}{-23.5$\pm$1.2} && \multicolumn{2}{c}{-5.4$\pm$2.1}	 && \multicolumn{2}{c}{-5.5$\pm$1.2}  && \multicolumn{2}{c}{-5.9$\pm$1.8}  \\
{H$\beta$ Abs EW}	& \multicolumn{2}{c}{9.8$\pm$0.4}   && \multicolumn{2}{c}{7.5$\pm$0.6}	 && \multicolumn{2}{c}{9.5$\pm$0.4}   && \multicolumn{2}{c}{7.2$\pm$0.6}   \\
{H$\alpha$ EW}		& \multicolumn{2}{c}{-89.5$\pm$2.9} && \multicolumn{2}{c}{-30.7$\pm$2.3} && \multicolumn{2}{c}{-27.9$\pm$0.9} && \multicolumn{2}{c}{-28.8$\pm$2.1} \\
{E(B-V)}                & \multicolumn{2}{c}{0.37$\pm$0.02} && \multicolumn{2}{c}{0.28$\pm$0.01} && \multicolumn{2}{c}{0.64$\pm$0.03} && \multicolumn{2}{c}{0.36$\pm$0.02} \\
\hline \\
{Galactic Lat.}         & \multicolumn{2}{c}{-53.626}       && \multicolumn{2}{c}{-52.102}       && \multicolumn{2}{c}{31.400}        &&  \multicolumn{2}{c}{31.351}       \\
{Galactic E(B-V)}       & \multicolumn{2}{c}{0.031}         && \multicolumn{2}{c}{0.029}         && \multicolumn{2}{c}{0.080}         && \multicolumn{2}{c}{0.070}         \\
\enddata
\tablecomments{
The spectra were de-reddened assuming \citet{storey87} case B for T$_{e}=1.25\times 10^{4}$ K and 
n$_{e}=$ 100 cm$^{-3}$ and the \citet{cardelli89} reddening law with $R_{\mbox{v}}=3.1$. 
For each galaxy, the measured flux ratio is given by F($\lambda$)/F(H$\beta$),
and the dereddened flux ratio by I($\lambda$)/I(H$\beta$).
The Galactic $E(B-V)$ is taken from \citet{schlegel98}.
The equivalent widths measure the H$\beta$ and H$\alpha$ Balmer emission lines using SPLOT, 
and the broad H$\beta$ absorption features using PAN. EWs are given in units of \AA.}
\tablenotetext{1}{The H$\beta$ flux is given for reference, with units of $10^{-17}$ erg s$^{-1}$ cm$^{-2}$.} 
\tablenotetext{2}{Equivalent width errors calculated from \citet{vollmann06}.}
\end{deluxetable}


\section{COMPARISON OF SAMPLE GALAXIES TO THE SDSS}\label{sec:SDSS}

\begin{figure}
  \begin{center}
    \plotone{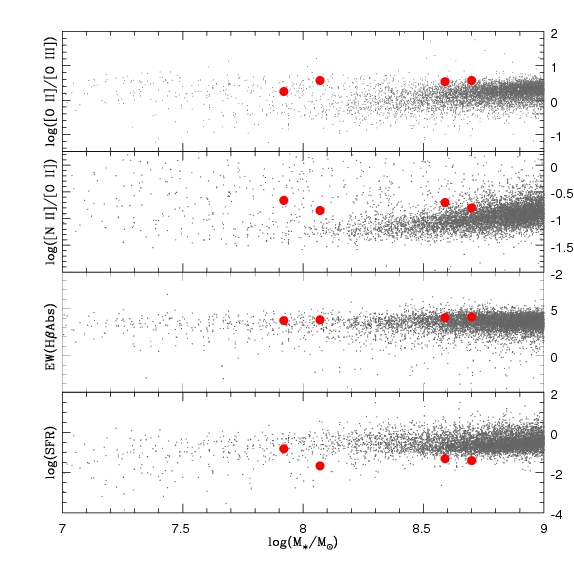}
      \caption{Properties of low-mass star-forming galaxies in the SDSS (gray dots) plotted against stellar mass.
      The four objects presented in this paper are shown as red dots in comparison.
	Interestingly, the [\ion{N}{2}]/[\ion{O}{2}] ratios are well above average for the mass 
	range, and the [\ion{O}{2}]/[\ion{O}{3}] ratios show that these objects have 
	significantly lower excitation than is typical.
	Additionally, their star formation rates are relatively low and their 
	H$\beta$ absorption equivalent widths are on the high end of the SDSS range.}
    \label{fig:6}
  \end{center}
\end{figure}

Our new observations allow us to examine these \citet{peeples08} galaxies 
against larger samples drawn from the MPA-JHU SDSS database.  
Since the ``very low mass" sample lacked [\ion{O}{2}] measurements, we are 
particularly interested in how the strengths of these emission lines compare.
Figure~\ref{fig:6} illustrates properties of the four observed galaxies with 
respect to star-forming galaxies in the SDSS that fall in the ``very low mass" 
sample range of $7 \le$ log(M$_{\star}$/M$_{\odot}) \le 9$.
We have further restricted the SDSS comparison sample to those objects with
s/n $\ge$ 5 in the relevant lines.

The top panel in Figure~\ref{fig:6} shows a comparison of the excitation,
measured by the [\ion{O}{2}]/[\ion{O}{3}] ratio.  The two more massive
galaxies are clearly discrepantly strong in [\ion{O}{2}] emission; these are
very low excitation galaxies.  
With decreasing mass, the SDSS comparison sample becomes relatively sparse, 
but the two lower mass galaxies have comparatively low excitation.   

The second panel shows a comparison of the
[\ion{N}{2}]/[\ion{O}{2}] ratio, primarily a function of the N/O abundance.
Again, for the two more massive galaxies, there is a clear offset from the
locus defined by the SDSS galaxies.  While the [\ion{N}{2}]/[\ion{O}{2}] ratios
are of comparable strength in the two lower mass galaxies, the trend in the
SDSS sample is less clear.  Extrapolating from the higher mass galaxies, the
lower mass galaxies would be clearly discrepant, but they lie in the middle of a
very sparse scatter in the diagram.
Quantitatively, in a comparison with a least-squares fit to the SDSS 
compilation, the values of log(N/O) for our four observed galaxies 
are $0.2-0.4$ dex higher than is typical for objects in the same 
stellar mass range.  

The lower two panels compare the equivalent widths of the underlying stellar
absorption and the star formation rates derived from H$\alpha$.
The four objects presented in this paper have H$\beta$ absorption equivalent 
widths on the high end of the SDSS distribution and star formation rates 
on the low side of the distribution.

In sum, the four galaxies which we have observed are outliers in a number of 
properties.  Given that oxygen abundances derived from strong emission lines
are subject to a number of systematic uncertainties, it is clearly warranted
to revisit their status as oxygen abundance outliers.


\section{OXYGEN ABUNDANCE DETERMINATIONS}\label{sec:metallicity}

Relative to the SDSS spectra, our MMT spectra have the advantages of higher 
signal to noise and the inclusion of the blue [\ion{O}{2}] $\lambda$3727 line. 
\citet{kniazev03} showed from SDSS spectra that O$^+$/H$^+$ ionic abundances 
can be determined reasonably well by observing the red [\ion{O}{2}] 
$\lambda\lambda$7320,7330 lines (as a substitute for [\ion{O}{2}] $\lambda$3727); 
however, because they are auroral lines, their strong sensitivity to temperature 
can result in relatively high abundance uncertainties. 
In all four MMT spectra (Figures \ref{fig:1}-\ref{fig:4}) the preferred [\ion{O}{2}] 
$\lambda$3727 emission line strengths are noticeably stronger than the 
[\ion{O}{3}] $\lambda$5007 lines, underscoring the importance of accurately accounting 
for the  contribution from the lower ionization state in determining oxygen abundances.

Accurate ``direct" oxygen abundance determinations from \ion{H}{2} regions 
require a measurement of the electron temperature (typically via observation 
of the temperature sensitive [\ion{O}{3}] $\lambda$4363 auroral line). 
However, as metallicity increases, cooling via metal lines becomes more 
efficient and the electron temperature decreases, making these intrinsically 
faint lines even more difficult to detect.  
Since we did not reliably detect [\ion{O}{3}] $\lambda$4363 in any of our targets, 
abundances must be estimated empirically or theoretically using 
relationships dependent upon relative strong line flux ratios.  

\citet{alloin79} and \citet{pagel79} were the first to provide strong-line calibrations,
where oxygen abundance is related to one or more ratios of recombination and
collisionally excited lines. 
Several other methods have since been developed and categorized as 
semi-empirical, empirical, or theoretical.
Semi-empirical calibrations were determined using a combination 
of electron temperature measurements at low metallicity and photoionization 
models at high metallicity in correspondence with observational limitations.
Empirical calibrations result from observations of \ion{H}{2} 
regions with electron temperature measurements.
However, the relatively small number of direct oxygen abundance determinations 
available are typically biased in the sense that they are based upon high-excitation
\ion{H}{2} regions only. 
In contrast, theoretical strong-line calibrations use \textit{ab initio} photoionization models. 
One advantage to theoretical models is that they allow a wide range in 
ionization parameter in addition to input metallicity. 
However, these models rely on simplified assumptions of nebular properties and so 
do not yet provide entirely realistic representations of \ion{H}{2} regions.
Most troubling is that all theoretical strong line abundance determination methods over-predict
abundances in the metal-rich (roughly solar metallicity and above) regime when 
compared to abundances determined from direct temperature measurements.

Here we review the original oxygen abundances as determined by T04 using Bayesian models, 
the revised strong line oxygen abundances calculated by P08, strong line oxygen 
abundances from our new MMT spectra (with and without the addition 
of the blue [\ion{O}{2}] emission), and the N/O relative abundance ratios. 
T04 found this type of comparison is valid for SDSS data by showing that  
analytic R23-metallicity relations roughly bracket the range of 
metallicities that they derive, concluding that their M-Z relationship is in 
line with previous strong-line calibrations.
The culmination of these efforts is presented in Figure~\ref{fig:7}, which
allows us to compare all of the oxygen abundance estimates with those 
expected from the M-Z relationship.
Note that Figure~\ref{fig:7} is shown for illustrative purposes only, and not to 
suggest which calibrator is more fundamentally correct.


\subsection{The T04 Oxygen Abundances}

P08 initially identified outliers from luminosities, gas-phase oxygen 
abundances, and stellar masses provided by T04.  
The T04 abundance calculations are based on a Bayesian statistical 
analysis of the strongest six emission lines using stellar population 
synthesis models from \citet{bruzual03} and photoionization models 
from CLOUDY \citep{ferland98}.
Note that while the T04 abundances were used to identify the outliers 
(in a self-consistent manner), P08 derived their own abundances, which 
are systematically lower than the T04 abundances. 
\citet[hereafter Y07]{yin07} compared the oxygen abundances from T04 
with oxygen abundances derived from direct temperature 
measurements (using auroral lines) and found the T04 oxygen 
abundances to be systematically higher on average.
Y07 further showed that the magnitude of this offset correlates 
well with the N/O abundance ratio, and concluded that the offset is due 
to the assumption of a single N enrichment trend in the underlying 
\citet{charlot01} modeling. 

\citet[hereafter KE08]{kewley08} compared the M-Z relationships 
derived from ten different strong-line methods and found large 
systematic discrepancies between empirical and theoretical calibrations.
In the KE08 study, T04 metallicities tend to be in the middle of the range 
when low metallicities are considered (12 + log(O/H) $\approx$ 8.5) and 
at the high end of the range when the high metallicities are considered 
(12 + log(O/H) $\approx$ 9.0). 
KE08 provide relations for converting one metallicity scale to another, but the 
results of the Y07 analysis imply that a simple conversion is not always sufficient.  
That is, not only can the absolute abundances provided by the strong-line 
methods be systematically offset from the true nebular abundances, but, 
in some cases, the strong-line methods do not even accurately rank the abundances
(see Table~\ref{tbl:3} and Figure~\ref{fig:7} for different rankings among the 4 objects presented here).


\subsection{The P08 Oxygen Abundances}

P08 derived their own oxygen abundances for the outliers 
identified from the T04 sample.  
Since there was insufficient information to reproduce the results of the Bayesian 
statistical analysis of T04, P08 used two of the methods investigated by KE08.
They used the [\ion{N}{2}]/[\ion{O}{2}] ratio as calibrated by KE08 for the 
``main" sample (where [\ion{O}{2}] $\lambda$3727 was observed),  
and the [\ion{O}{3}]/[\ion{N}{2}] ratio (the ``O3N2'' method) as calibrated by 
\citet{pettini04} for the remaining galaxies in the ``very low mass" sample.  
P08 did not use the $R_{23}$ diagnostic for their ``main" sample because of concerns 
regarding its calibration at high metallicities (the presumed regime for their samples).
P08 found that their oxygen abundances derived from the [\ion{N}{2}]/[\ion{O}{2}] 
ratio for the ``main" sample agreed well with the T04 oxygen abundances.  
However, the O3N2 abundances were consistently lower than the T04 abundances 
by roughly 0.3 dex (in concordance with the offsets determined by KE08).  
The net effect is that the abundances derived by P08 are lower than those derived by T04.
P08 concluded that, regardless of the metallicity calibration used, their sample of 41 
galaxies are true high-metallicity, low mass outliers from the mass-metallicity relation. 
Despite any offsets between the T04 and P08 data sets, both use a strong line calibration 
based partially on [\ion{N}{2}] strength, which make abundance estimates liable to 
overestimates if the galaxies are nitrogen-enhanced (see \S~\ref{sec:N2} and \S~\ref{sec:N/O}).


\subsection{Strong Line Methods Revisited}

We have chosen to re-compute the oxygen abundances for this sample using the MMT 
spectra with several strong line calibrations: the O3N2 method, the N2 method, the 
R$_{23}$ index, and determinations using assumed temperatures. 
This serves to highlight differences amongst methods, outline the resulting range of possible 
abundances, and demonstrate that strong line calibrations are not appropriate for the present sample.
These measurements are discussed below and compared in Figure~\ref{fig:7}.


\subsubsection{``O3N2" Method}\label{sec:O3N2}
The O3N2 method is one of two empirical calibrations used by P08 for 
the ``main" sample, and the only calibration used for the ``very low mass"
sample since these objects lacked the necessary [\ion{O}{2}] measurements.
It was introduced by \citet[hereafter PP04]{pettini04} using empirical fits 
to strong line ratios from \ion{H}{2} regions with ``direct" oxygen abundances. 
Derived for the purpose of measuring metallicities in galaxies at high redshift, 
lines close in wavelength are used to mitigate the need for flux calibrations 
and reddening corrections. 
Direct metallicities were compared to the ratio of ([\ion{O}{3}] 
$\lambda$5007/H$\beta$)/([\ion{N}{2}] $\lambda$6584/H$\alpha$) 
for a sample of 137 \ion{H}{2} regions.
See the resulting O3N2 relationship of Equation~3 in PP04,
where O3N2 = $\log$(([\ion{O}{3}] $\lambda5007/$H$\beta$)/([\ion{N}{2}] 
$\lambda6584/\mathrm{H}\alpha))$. 

P08 favored the O3N2 method with its high sensitivity to oxygen abundance.
Indeed, O3N2 is a good calibration in the high-metallicity regime, where [\ion{N}{2}] tends to 
saturate, but the strength of [\ion{O}{3}] continues to decrease with increasing metallicity.
Note that the choice by PP04 to use only objects with direct metallicities, 
in an effort to provide a more secure calibration, introduces a bias because 
low excitation spectra (i.e., relatively low values of $\lambda$5007/$\lambda$3727) 
are excluded from the calibration. 
Thus, the unintended consequence of the choice to limit the sample to
objects which were perceived to have more accurate abundances has
resulted in a biased sample.  This biased sample gives the impression
of a smaller scatter in the relationship than occurs in nature \citep[cf.,][]{yin07}.
  
To confirm consistency with P08, we duplicated the O3N2 measurements for 
the SDSS spectra and calculated O3N2 from our MMT spectra. 
The results are listed in Table \ref{tbl:3} and closely match the measurements 
by P08 using the same O3N2 methodology; they are lower than the values 
found by T04 by 0.2 - 0.3 dex.

Y07 showed a large scatter in the comparison of abundances derived from 
the O3N2 calibration with oxygen abundances derived from the direct method.  
They suggest that the scatter could be due to excluding the ionization 
parameter in the O3N2 calibration. 
The importance of the ionization parameter strengthens in the low-metallicity 
regime where O3N2 is much less dependent on metallicity. 
Additionally, at low Z, N is thought to be a primary element, implying that the 
N/O ratio is independent of O/H.

Since the calibration of the O3N2 method does not account for possible 
variations in N/O for a given O/H, inherent biases are possible.
\citet[hereafter PMC09]{perez-montero09} found that a strong correlation 
exists between metallicity derived from the O3N2 parameter and the N/O ratio, 
such that when the N/O is enhanced, the O3N2 
strong line calibration tends to overestimate the O/H abundances. 
PMC09 corrected for this dependence by using 12 + log(O/H) versus O3N2 
residuals to modify the O3N2 calibration (see Equation~8 in PMC09).
Using the N/O calculations from \S~\ref{sec:N/O} and given in Table~\ref{tbl:4},
we calculated the revised O3N2 abundances and list them in Table \ref{tbl:3}.
This calibration lowers the O3N2 abundances only slightly (by
an average of $\sim0.1$ dex) relative to PP04 O3N2 estimates.
In \S~\ref{sec:expected} we look at the higher abundances from the O3N2 method 
relative to 
relationships predicted by direct and photoionization model abundances.


\subsubsection{``N2" Indicator}\label{sec:N2}

\citet[hereafter D02]{denicolo02} proposed the use of the [\ion{N}{2}] 
$\lambda$6584/H$\alpha$ ratio as a sensitive metallicity indicator. 
This is a promising estimator: like O3N2 it eliminates uncertainties 
due to reddening and flux calibrations, and functions well for metallicity 
values below the [\ion{N}{2}] saturation level. 
The original N2 calibration developed by D02 (see Equation~2 in D02), 
where N2 = $\log$([\ion{N}{2}] $\lambda6584/\mathrm{H}\alpha)$,
was based on $\sim$155 \ion{H}{2} regions and used a least squares 
fit with an estimated uncertainty of $\sim$0.2 dex. 
For our objects, this method produces oxygen abundances 
(8.61 $\le$ 12 + log(O/H) $\le$ 8.70) in agreement with the P08 values.

PP04 re-calibrated this parameter, primarily with electron temperature-based 
metallicities, characterizing the fit both linearly and with a third-order polynomial. 
The polynomial fit is given in Equation~2 of PP04, 
yielding oxygen abundances of 8.45 $\le$ 12 + log(O/H) $\le$ 8.56.
The results for both calibrations are listed in Table~\ref{tbl:3}. 

Note, again, as with the PP04 calibration of the O3N2 method, the 
exclusive use of calibrators determined with measured electron 
temperatures underestimates the scatter in the relationship, and 
biases against low-ionization \ion{H}{2} regions. 
\citet{yin07} found similar results in their analysis of the N2 method 
as found for the O3N2 method. The uncertain roles of the ionization 
parameter and the relationship 
between N/O and O/H as a function of metallicity result in appreciable 
uncertainty and scatter in the N2 method.
Expanding on this idea, PMC09 found an expectedly strong dependence of  
metallicities predicted by N2 on the N/O ratio.
PMC09 derived a calibration correcting for this effect (see their Equation~13).
Using N/O ratios from \S~\ref{sec:N/O} and tabulated in Table~\ref{tbl:4},
we present the results of this calibration in Table~\ref{tbl:3}.
This method predicts oxygen abundances which are $\sim0.1$ and $\sim0.2$ dex smaller 
on average than those predicted by the PP04 and D02 N2 methods respectively.
Shifts to smaller oxygen abundances of this size can account for roughly half of the
offset from the L-Z and M-Z relationships.


\subsubsection{R$_{23}$ Index}\label{sec:R23}

\citet{pagel79} promoted the use of the R$_{23}$ index, 
R$_{23}$ = ([\ion{O}{2}] $\lambda3727$ + [\ion{O}{3}] $\lambda\lambda$4959, 5007)/(H$\beta$),
as a good estimate of oxygen abundance in the absence of an electron temperature measurement. 
Because the optical [\ion{O}{2}] and [\ion{O}{3}] emission lines decrease at 
both high abundances (due to an increasing role of fine structure line cooling) 
and low abundances (due to the decrease in the relative number of oxygen 
atoms), the relationship is bi-valued and, therefore, potentially ambiguous. 
The turn-around in the relationship occurs at oxygen abundances near 12 + log(O/H) 
$\approx$ 8.4 where R$_{23}$ reaches a maximum of $\approx$10. 
However, the degeneracy between the two branches of solutions can, in
most situations, be broken with an additional determinant. 
Also, since R$_{23}$ is based on emission lines with significant separation 
in wavelength, accurate reddening corrections and uncertainties are important. 
Below, we discuss two R$_{23}$ calibrations, one constructed from photoionization 
models, and another empirically based.  
Since P08 believed the sample to be high-metallicity, they did not consider the 
R$_{23}$ diagnostic due to the potential of [\ion{O}{2}] + [\ion{O}{3}] to saturate 
and the lack of [\ion{O}{2}] measurements in their ``very low mass" sample.


\citet{mcgaugh91} created a calibration based on theoretical photoionization 
models using R$_{23}$ and the additional O$_{32}$ index:
O$_{32}$ = ([\ion{O}{3}] $\lambda\lambda4959,5007$)/([\ion{O}{2}] $\lambda3727$).
McGaugh developed this model using the photoionization code CLOUDY 
\citep{ferland98} and zero-age \ion{H}{2} region models, 
accounting for photoionization parameter variations. 
To discriminate between the two branches, \citet{vanzee98}, 
followed by others, advised using the ratio of 
([\ion{N}{2}] $\lambda$6584)/([\ion{O}{2}] $\lambda$3727). 
\citet{mcgaugh94} suggested that [\ion{N}{2}]/[\ion{O}{2}] is approximately 
$<$ 0.1 for low abundances and $>$ 0.1 for high abundances, giving a 
rough distinction between lower and upper branches.
Note, however, that \citet{mcgaugh94} also points out (his Figure 3) that
there is a dependence on the N/O ratio on the behavior of this discriminant.
For the four galaxies in our sample, [\ion{N}{2}]/[\ion{O}{2}] ranges from 
0.15 to 0.22 with an average uncertainty of $\sim$0.02, 
naively suggesting that they are upper branch objects near the turn-around region. 
We determined oxygen abundances using the analytic equations of the 
semi-empirical calibration from \citet{kobulnicky99}, 
which have an estimated accuracy of $\sim$0.15 dex \citep{kobulnicky04}.
The results for both upper and lower branch calculations are listed in Table~\ref{tbl:3}. 
The upper branch values define an upper limit to the metallicity estimates 
and are roughly equal to the P08 values, 
while the lower branch values are approximately 0.3 to 0.5 dex lower and 
would be consistent with the scatter in the luminosity-metallicity relation.


\citet{pilyugin01a} and \citet{pilyugin01b} empirically calibrated the R$_{23}$ 
index with electron temperature based \ion{H}{2} region oxygen abundances. 
This relation was later refined by \citet[][hereafter PT05]{pilyugin05}, 
where the resulting fit incorporates the excitation parameter 
(which accounts for the effect of the ionization parameter) 
P = ([\ion{O}{3}] $\lambda\lambda${4959,5007}/H$\beta$)/R$_{23}$. 
The upper and lower branches correspond to electron temperature based 
metallicities with 12 + log(O/H) $>8.25$ and 12+ log(O/H) $<8.0$ respectively
(see Equations~22 and 24 in PT05).
The PT05 model upper and lower branch values, both of which are 
several tenths of dex below the P08 values and have an estimated 
accuracy of $\sim$0.1 dex, are given in Table~\ref{tbl:3}. 
However, note that all four of our objects lie outside of the calibrated 
region for the PT05 method (see PT05, Figure 12).  

The [\ion{O}{3}]/[\ion{O}{2}] ratio is sensitive to the ionization parameter, 
and important for characterizing the physical conditions in the \ion{H}{2} region. 
Since the age of an \ion{H}{2} region is linked to the evolution of the ionization 
parameter and shape of the ionizing spectrum, photoionization models which
assume a zero-age main sequence can lead to systematic errors.
\citet{vanzee06a} and \citet{vanzee06b} discuss these effects and suggest they 
may be significant for \ion{H}{2} regions with log(O$_{32}) < -0.4$, such as 
for three of the four objects in our sample (see Table~\ref{tbl:3} and \S~\ref{sec:discussion}).
The small $\lambda$5007/$\lambda$3727 ratios in these spectra are indicative 
of a low ionization parameter, and correspond to very small P values 
(see Table~\ref{tbl:3}) that are inconsistent with our high values of 
log(R$_{23}$) for the PT05 upper branch calibration.
Similarly, our low P values fall outside the range of calibration for their lower branch.
If we naively extrapolate their calibrations, these objects fall close to or within the 
turn-around region which has an abundance range of $8.0<$ 12 + log(O/H) $<8.4$. 

[\ion{N}{2}]/H$\alpha$ is often used as a second indicator of branch division. 
However, as it is less sensitive to metallicities and more responsive to ionization 
than [\ion{N}{2}]/[\ion{O}{2}], the division is less clearly defined. 
The latter ratio loses the ability to clearly break the degeneracy near the turn-around 
point of {[\ion{N}{2}]/[\ion{O}{2}]} $\sim{0.1}$ (or $12+\log(\mathrm{O/H})\sim{8.4}$). 
With its dependence on both oxygen abundance and excitation, R$_{23}$ seems like a
superior metallicity indicator, but the ambiguity between branches requiring a secondary measurement
like [\ion{N}{2}]/H$\alpha$ or [\ion{N}{2}]/[\ion{O}{2}] clearly render it less than ideal.
The enhanced [\ion{N}{2}]/[\ion{O}{2}] values for the four objects investigated in this 
paper result in highly uncertain R$_{23}$ metallicity estimates. 
Note, however, that the turn-around between the two branches happens at relatively 
low oxygen abundances, for all values of excitation, already indicating that the higher 
oxygen abundances reported by T04 and P08 are less likely. 

Finally, we note the possible utility of the red [\ion{O}{2}] emission 
lines in low redshift SDSS spectra.
\citet{kniazev03} showed that SDSS abundances estimated using the red 
[\ion{O}{2}] $\lambda\lambda$7320,7330 lines as a substitute for 
[\ion{O}{2}] $\lambda$3727 yield comparable results for spectra with 
direct electron temperature measurements. 
Since the red [\ion{O}{2}] emission lines were detected in the SDSS-022628 
spectrum (Figure~\ref{fig:1}) we were able to derive alternative R$_{23}$ 
abundance estimates for that galaxy.  By using the relative emissivities to
estimate the relative strengths of the blue [\ion{O}{2}] emission lines 
from the red [\ion{O}{2}] emission lines, we found that the 
$\lambda$5007/$\lambda$3727 ratio is less than unity (suggesting relatively low 
ionization) over the range $1\times10^{4}$ K $\le$ T$_{e}$ $\le$ 
$1.25\times10^{4}$ K (for n$_{e}=100$ cm$^{-3}$).
The corresponding R$_{23}$ calibrated O/H values are consistent with our MMT 
spectra findings, but, of course, the branch choice ambiguity remains.


\begin{deluxetable}{cccccc}
\tabletypesize{\footnotesize}
\tablecaption{Metallicity Estimates for \ion{H}{2} Regions Using Empirical Methods \label{tbl:3} }
\tablewidth{0pt}
\tablehead{
& \multicolumn{5}{c}{MMT Data} \\
\cline{2-6}
\colhead{Quantity} & \colhead{SDSS-022628} & \colhead{SDSS-024121} & \colhead{SDSS-082639} & \colhead{SDSS-082633} }
\startdata
{O3N2}						& 0.77$\pm$0.05	 & 0.45$\pm$0.08  & 0.28$\pm$0.11  & 0.66$\pm$0.13  \\
{N2}						& -0.71$\pm$0.04 & -0.63$\pm$0.06 & -0.58$\pm$0.06 & -0.69$\pm$0.09 \\
{$\log$ R$_{23}$}				& 0.60$\pm$0.02	 & 0.71$\pm$0.04  & 0.64$\pm$0.06  & 0.71$\pm$0.05  \\
{$\log$ O$_{32}$}				& -0.20$\pm$0.04 & -0.64$\pm$0.08 & -0.74$\pm$0.12 & -0.50$\pm$0.11 \\
{[\ion{N}{2}]$/$[\ion{O}{2}]}			& 0.22$\pm$0.01	 & 0.16$\pm$0.01  & 0.20$\pm$0.02  & 0.15$\pm$0.01  \\
{[\ion{N}{2}]$/$H$\alpha$}			& 0.20$\pm$0.01	 & 0.24$\pm$0.01  & 0.26$\pm$0.02  & 0.20$\pm$0.02  \\
{P}						& 0.39$\pm$0.01	 & 0.19$\pm$0.01  & 0.15$\pm$0.02  & 0.24$\pm$0.03  \\
{N/O}						& 0.16$\pm$.01	 & 0.12$\pm$0.01  & 0.15$\pm$0.02  & 0.11$\pm$0.01 \\
\hline
{Pettini and Pagel (O3N2)}			&	8.48	&	8.59	&	8.64	&	8.52	\\
{P{\'e}rez-Montero and Contini (O3N2)}		&	8.37 	&	8.52 	&       8.54 	&       8.47 	\\
{Pettini and Pagel (N2)}			&	8.45	&	8.51	&	8.56	&	8.46	\\
{Denicolo (N2)}					&	8.60	&	8.66	&	8.70	&	8.61	\\
{P{\'e}rez-Montero and Contini (N2)}		&	8.30 	&	8.44 	&       8.42 	&       8.41 	\\
{Kewley and Dopita (\ion{N}{2}$/$\ion{O}{2})}	&	8.83	&	8.74	&	8.80	&	8.71	\\
{McGaugh Upper (R$_{23}$)}			&	8.77	&	8.63	&	8.70	&	8.64	\\
{McGaugh Lower (R$_{23}$)}			&	7.81	&	8.13	&	8.06	&	8.08	\\
{Pilyugin Upper}				&	8.40	&	8.03	&	8.07	&	8.11	\\
{Pilyugin Low}					&	7.58	&	7.43	&	7.22	&	7.55	\\
{van Zee expected from M$_{B}$}\tablenotemark{1} &	8.22	&	8.17	&	8.07	&	7.97	\\
\hline 
&&&&&
					   					                                                                		 		\\
& \multicolumn{5}{c}{SDSS Data}       					                                \\
\cline{2-6}
{}	    	                        & {SDSS-022628} & {SDSS-024121} & {SDSS-082639} & {SDSS-082633} \\
\hline
{Pettini and Pagel (O3N2)}                  	& 8.53        & 8.61        & 8.68        & 8.50        \\
{Pettini and Pagel (N2)}   		        & 8.48        & 8.54        & 8.59        & 8.43        \\ 
{Denicolo (N2)}		   		        & 8.63        & 8.69        & 8.73        & 8.58        \\
{Tremonti (T04)}                                & 8.82        & 8.90        & 8.86        & 8.69        \\
{Peeples 08}                                    & 8.51        & 8.61        & 8.66        & 8.49	\\    
\enddata
\tablecomments{Values for various methods of determining metallicity estimates from the strong lines of the MMT spectra are listed, 
with the O3N2 and N2 estimates for the SDSS spectra given below for comparison. 
Interestingly, the N2 method produces similar values for all 4 galaxies, and the O3N2 method reproduces similar values to those found by P08.} 
\tablenotetext{1}{Metallicity estimates calculated using Equation~\ref{eq:vanzee} of this paper, as determined by \citet{vanzee06b}.}

\end{deluxetable}


\begin{figure}
  \begin{center}
      \subfigure[Luminosity-Metallicity Relation]{\includegraphics[scale=0.4]{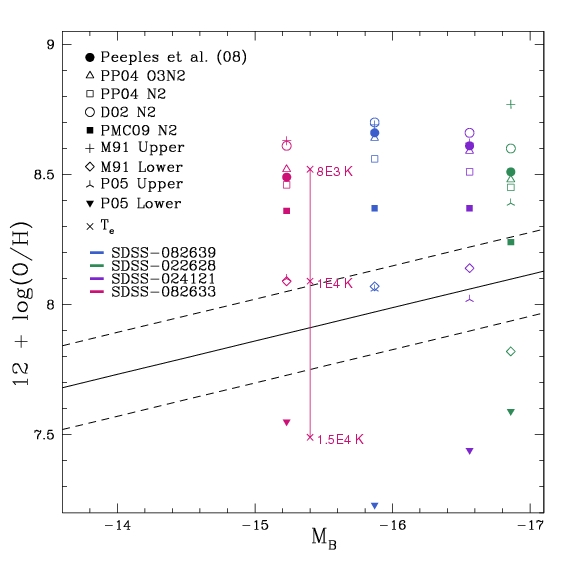}}
          \subfigure[Mass-Metallicity Relation]{\includegraphics[scale=0.4]{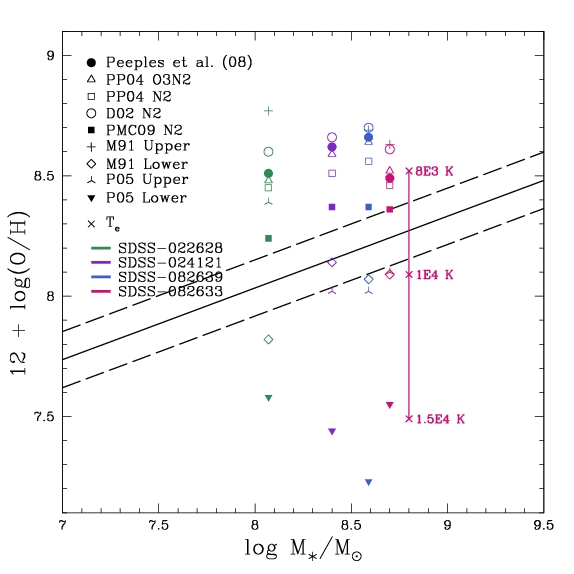}}
	      \caption{\small The figure (a) shows $12+\log(\mbox{O/H})$ vs. M$_{B}$ for our sample. 
	The solid line is the linear fit to dwarf galaxies found by \citet{lee06b} for comparison, 
	with the dashed lines representing the 1 $\sigma$ uncertainty.
	This relationship is similar to that of \citet{vanzee06b}, 
	but provides an average dispersion for both the L-Z and M-Z relationships.
	The points plotted display calculated oxygen abundances for our sample. 
	Each method is denoted with a different symbol, where objects are differentiated 
	by color: green for SDSS-022628, purple for SDSS-024121, blue for 
	SDSS-082639, and pink for SDSS-082633. 
	The `x' symbols denote the oxygen abundances calculated from emissivities 
	assuming various electron temperatures for SDSS-082633. 
	Figure (b) displays $12+\log(\mbox{O/H})$ vs. $\log(\mbox{M}_{\star}/\mbox{M}_{\odot})$. 
	Again, the solid line depicts the linear relationship and 1 $\sigma$ uncertainty of \citet{lee06b} for mass.}
       \label{fig:7}
   \end{center}
\end{figure}


\subsection{N/O Relative Abundances}
\label{sec:N/O}
Since many of the methods of abundance determination discussed here rely on 
N abundance values either to calibrate O/H values or to discriminate between 
bi-valued solutions, a confident measurement of the N abundance is important.  
Due to the relative insensitivity of the derived N/O ratio to electron temperature, 
it is possible to get a reliable estimate of this ratio in the absence 
of an electron temperature determination.  
We calculated the N/O values for our four galaxies (see Table~\ref{tbl:4}), 
assuming that  N$^{+}$/O$^{+}$ is approximately equivalent to that of N/O 
\citep[based on their similar ratio of ionization potentials,][]{vce93} 
and an electron temperature of 12,500 $\pm$ 2,500 K.  

The N/O versus O/H trend is well studied in galaxies of varying types.
\citet{vce93} presented a thorough overview of theoretical 
expectations and observations available at the time.
A salient point is that N can be produced as both a primary and a secondary 
element and that the secondary component is expected to be delayed relative 
to oxygen and to dominate at high abundances.
\citet{garnett90} convincingly demonstrated that there can be significant 
scatter in N/O at a given O/H, and \citet{izotov99} confirmed that this 
scatter is significant for oxygen abundances above 12 + log(O/H) = 7.7.
\citet{garnett90} proposed that much of the scatter could be explained by 
the time delay between producing oxygen and secondary nitrogen.
However, \citet{henry06} conclude that the scatter in N/O could be produced
by a variety of causes.

As already indicated in \S~\ref{sec:SDSS}, the four objects in our sample 
have unusually high values of N/O.
PMC09 and, later, \citet{amorin10} (in their study of the abundances of 
the ``green pea'' galaxies)
present a modern view of N/O versus O/H assembled from a sample of 475 
\ion{H}{2} objects and the SDSS DR7 catalog of observations respectively.
A clear trend is obvious, but with scatter in log(N/O) on the order of 
0.5 dex or more for a given O/H.  
An inspection of Figure~2 from \citet{amorin10} shows that our four galaxies, 
which range between $-0.8 \le$ log(N/O) $\le -1.0$, 
correspond to values of 12 + log(O/H) $\approx$ 8.6 (derived from following 
the ridgeline of the data). 
Thus, empirically calibrated strong-line methods, such as O3N2 and N2, 
would predict oxygen abundances close to this value. 
However, the galaxies in this range of log(N/O) also extend to abundances 
as low as 12 + log(O/H) $\approx$ 7.9.
From a plot of log(N/O) versus stellar mass for the SDSS compilation 
given by \citet{amorin10} (their Figure~3), our values of log(N/O) are 
roughly 0.6 dex higher than is typical for objects in the same stellar 
mass range. 


\subsection{Expected Oxygen Abundances}\label{sec:expected}

All strong-line calibrations correspond to an assumed electron temperature, 
so it is informative to explore the affects of varying this parameter.
Since our MMT spectra include emission lines from both O$^+$ and O$^{++}$,
we can calculate an expected range in O/H values based on an assumed 
reasonable possible range in electron temperature.
As an example, oxygen abundances were calculated for SDSS-082633 assuming
electron temperatures of T$_{e}=8\times10^{3}$ K, $1\times10^{4}$ K, and 
$1.5\times10^{4}$ K.  The resulting values of oxygen abundance cover the
range of $12 + \log(\mathrm{O/H}) = 7.5 - 9$, and are indicated by
`x' symbols in Figure~\ref{fig:7}; similar results are found for
the other 3 galaxies.
Thus, we cannot rule out any of the various strong line calibration 
estimates based on this consideration.

Finally, we can estimate the expected oxygen abundances from the
luminosity-metallicity relationship and compare these to the strong line
estimates in Figure~\ref{fig:7}.
Each calibration is represented by a different symbol, and each of our four galaxies 
is designated with a different color.
Using the M-Z relationship of \citet{vanzee06b}\footnote{\citet{vanzee06b} used both direct and 
empirical oxygen abundances from the literature for 50 objects in determining their M-Z relationship.},
\begin{equation}
	12 + \log(\mbox{O/H}) = 5.67 - 0.151\ \mbox{M}_{B},
	\label{eq:vanzee}
\end{equation}
we calculate expected oxygen abundances of 8.22, 8.17, 8.07, and 
7.97 for the four objects in our sample. 
The M-Z relationship found by \citet{lee06b}, based on oxygen abundances obtained 
via the direct method and corroborated by \citet{marble10}, covers the relevant range
in luminosity for our sample and is also plotted in Figure \ref{fig:7}.
(Note that the range of the T04 relation does not extend down to encompass our 
low-luminosity sample). 
For each galaxy, a large spread is seen amongst the various indicators, 
highlighting the large uncertainties inherent in metallicity determinations.  


\begin{deluxetable}{ccccc}
\tabletypesize{\footnotesize}
\tablecaption{Estimates of the N/O Ratios for Our Sample \label{tbl:4} }
\tablewidth{0pt}
\tablehead{
& \multicolumn{4}{c}{log(N/O)} \\
\cline{2-5}
& \multicolumn{4}{c}{MMT Data} \\
\colhead{Temperature} & \colhead{SDSS-022628} & \colhead{SDSS-024121} & \colhead{SDSS-082639} & \colhead{SDSS-082633} }
\startdata
{$1.00\times10^{4}$ K} & -0.94		  & -1.08	     & -0.98		& -1.12	 		\\
{$1.25\times10^{4}$ K} & -0.80	          & -0.94	     & -0.84		& -0.98       		\\
{$1.50\times10^{4}$ K} & -0.70	          & -0.84	     & -0.74		& -0.88       		\\
{Adopted}              & -0.80 $\pm$ 0.12 & -0.94 $\pm$ 0.12 & -0.84 $\pm$ 0.12 & -0.98 $\pm$ 0.12 	\\
\enddata
\tablecomments{N/O values for our four galaxy sample, calculated for a range of reasonable electron temperatures (12,500 $\pm$ 2,500 K).} 
\end{deluxetable}


\section{DISCUSSION}\label{sec:discussion}
\subsection{Oxygen Abundances from Direct Comparison with Photoionization Models}\label{sec:comparison}

The MMT spectra show that three of the objects observed have discrepantly 
low excitation (i.e., low $\lambda$5007/$\lambda$3727 ratio;  log(O$_{32}) < -0.4$).  
This implies that the observed nebulae may have relatively low 
ionization parameters and/or be excited by an aging stellar 
population, which explains why we do not see other ions with high 
ionization potentials (such as \ion{He}{2} $\lambda$4686) in our spectra. 
Since most strong line methods are calibrated for zero-aged main sequence 
starbursts, and many do not incorporate the effects of excitation, 
one must exercise care in applying these methods to determine 
oxygen abundances \citep[see discussion in][]{vanzee06b}.
In general, the strong line methods tend to over-estimate the oxygen 
abundance for low excitation nebulae, and this offset can be as 
severe as $\sim$0.6 dex \citep{vanzee06b}.
Note that it may be possible to mitigate some error by combining empirical 
and theoretical determinations using an average as \citet{moustakas10} suggest. 

Additionally, all four of the objects observed have relatively high 
values of N/O compared to systems of similar stellar mass. 
As discussed by \citet{yin07}, PMC09, and \citet{amorin10}, the assumption 
of a single relationship between N/O and O/H in photoionization model 
calibrations of the strong line methods can lead to discrepant results 
when the N abundance does not fit this assumed trend.
At higher than average values of N/O, the strong line methods tend to 
over-estimate the oxygen abundance.

\begin{figure}
  \begin{center}
    \subfigure[Oxygen Abundance Models]{\includegraphics[scale=0.4]{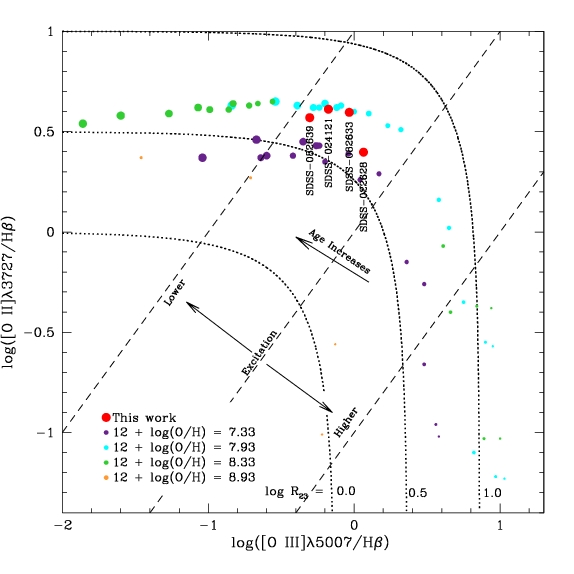}}
        \subfigure[Low-Mass SDSS Oxygen Abundances]{\includegraphics[scale=0.4]{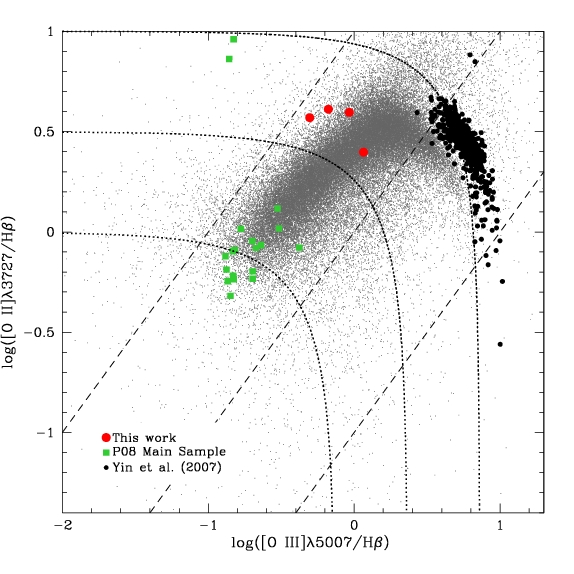}}
      \caption{(a) A comparison of the results from our MMT spectra with results from a large range 
      	of photoionization models evolved over 10 Myr as determined by \citet{stasinska96} 
	(symbol size corresponds to aging sequence). 
	The overlaid dotted contours represent lines of constant R$_{23}$; from left to right log(R$_{23}$) = 0.0, 0.5, and 1.0. 
	The turn-around region encompasses log(R$_{23}$) values of approximately 0.5 to 1.0.
	 Dashed lines display excitation, where constant excitation of [\ion{O}{3}]/[\ion{O}{2}]= 1 coincides with the central line,
	 [\ion{O}{3}]/[\ion{O}{2}]=0.1 is the low excitation line, and [\ion{O}{3}]/[\ion{O}{2}]=10 is the high excitation line.
	Our four spectra are all found near the maximum values of R$_{23}$ for relatively low 
	excitation nebulae, indicating probable 12 + log(O/H) values in the range of 7.9 to 8.3.
	(b) Same plot of [O III] versus [O II] as Figure (a).
	The numerous gray dots are the MPA-JHU data for SDSS galaxies in the range of 
	$7.0 \le$ log(M$_{\star}$/M$_{\odot}) \le 9.0$, which have higher excitation than 
	SDSS-024121, SDSS-082639, and SDSS-082633.
	Values presented by \citet{yin07} for a sample of SDSS galaxies containing direct 
	method abundances are also included for comparison.
	Notice that three of our objects from the ``very low mass" sample are discrepantly low 
	excitation relative to the SDSS and \citet{yin07} samples, whereas the ``main" sample 
	has ordinary excitation values for the most part.}
    \label{fig:8}
  \end{center}
\end{figure}

The combination of relatively low excitation and relatively high N/O 
values leads to a large bias in the results from the strong line calibrations.
In Figure~\ref{fig:8}(a) we have plotted a comparison of the [\ion{O}{2}] 
and [\ion{O}{3}] line strengths for our four objects to photoionization 
models from \citet{stasinska96}.
Each set of different colored points represents the evolution of a 
starburst assuming a Salpeter initial mass function (M$_{up}=100$ M$_{\odot}$), 
a stellar mass between $10^6$--$10^9$ M$_{\odot}$, 
and a metallicity between 12 + log(O/H) = 7.33 and 8.93. 
Increasing symbol size denotes a progression from 1--10 Myr. 
These models are particularly valuable because they demonstrate the 
additional scatter that can be introduced as the exciting stars of an 
\ion{H}{2} region age, and both the ionization parameter and shape of 
the ionizing spectrum evolve. 
Figure~\ref{fig:8}(a) shows that \ion{H}{2} region models with oxygen 
abundances in the range of 7.9 $\le$ 12 + log(O/H) $\le$ 8.3 tend 
to have maximum values of R$_{23}$ (thus defining the turn-around regime 
in R$_{23}$; recall log(R$_{23}$) has an upper limit near 1.0).  
Our four spectra fall among the points for maximum R$_{23}$ at lower excitation.
Figure~\ref{fig:8}(a) demonstrates that the four objects are inconsistent 
with relatively high (or relatively low) values of O/H, and, actually, 
are consistent with the values of O/H predicted from either the 
luminosity-metallicity relationship or the M-Z relationship.

Note that in constructing Figure~\ref{fig:8}(a), some of the \citet{stasinska96} 
high-metallicity models showed more  scatter in this diagram.  
These models included contributions by Wolf-Rayet (W-R) stars to the ionizing 
spectra, resulting in very hard radiation fields.  
We have not plotted these model points in Figure~\ref{fig:8}(a) because our non-detections 
of \ion{He}{2} $\lambda$4686 indicate very little contribution from W-R stars at present.

MPA-JHU data for SDSS galaxies in the mass range of $7 \le$ log(M$_{\star}$/M$_{\odot}) 
\le 9$ are plotted in Figure~\ref{fig:8}(b) (gray points). 
Three of the four objects in the present sample, SDSS-024121, SDSS-082639, and SDSS-082633, are 
discrepantly low excitation nebulae relative to the SDSS sample, which tend to have higher excitation.
However, most of the P08 ``main" sample objects have ordinary excitation values.
Note that seven of the 24 ``main" sample objects did not have z $>0.024$ (consequently, 
they didn't have [\ion{O}{2}] $\lambda3727$ in their spectra), and so could not be plotted here.

\begin{figure}
  \begin{center}
    \plotone{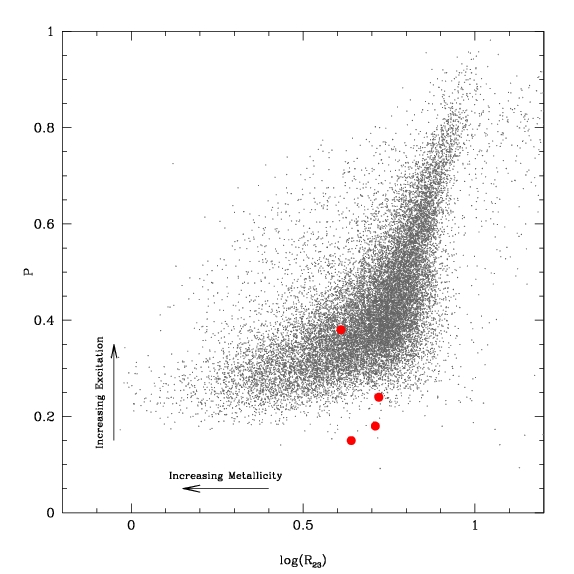}
      \caption{Excitation parameter P versus the metallicity sensitive R$_{23}$ index for 
      the four objects in this paper (red dots) and the low-mass star-forming galaxies in the SDSS (gray dots).
      Arrows indicate direction of increasing excitation and increasing metallicity for the R$_{23}$ upper branch. 
      For three of these objects, the excitation is rather low for their R$_{23}$ values.}
    \label{fig:9}
  \end{center}
\end{figure}

A complementary analysis is seen in Figure~\ref{fig:9} where excitation (P) 
is plotted against log(R$_{23}$) for this same set of low mass SDSS galaxies 
(also see \cite{moustakas10} Figure 12 for a similar comparison).
As discussed in \S~\ref{sec:R23}, three of our four objects lie well
below the mean excitation for a given R$_{23}$ value.
Figure~\ref{fig:9} also shows an upper limit of log(R$_{23}$) naturally falls 
near 1.0, where the present sample lies within the turn-around region 
defined by $0.5 \lesssim$ log(R$_{23}) \lesssim 1.0$, corresponding to 
7.9 $\le$ 12 + log(O/H) $\le$ 8.3.
A similar conclusion is drawn from the N2 calibration of PMC09 (\S~\ref{sec:N2}) 
which suggests an oxygen abundance range of 8.30 $\le$ 12 + log(O/H) $\le$ 8.44 
for the present objects based on the abnormal N/O variations present
($\sim0.1-0.2$ dex less than other strong-line N2 calibrations).
While we can't constrain the oxygen abundance more precisely for this sample (based 
on the arguments here in \S~\ref{sec:comparison}) than to put them in the range of 
7.9 $\le$ 12 + log(O/H) $\le$ 8.4, this is sufficient to disqualify them as high-metallicity M-Z outliers.

Note that the recalibration of the O3N2 and N2 abundance indicators by PMC09 reduced the inferred
oxygen abundances, but only by about half of what was needed to bring the objects in line with their 
expected abundances that we are proposing here. 
Even after recalibration, there is still a significant amount of scatter in these relationships.
We speculate that taking into consideration the additional concern of low excitation would move
these points further in the direction of concordance with expectations.


\begin{figure}
  \begin{center}
    \plotone{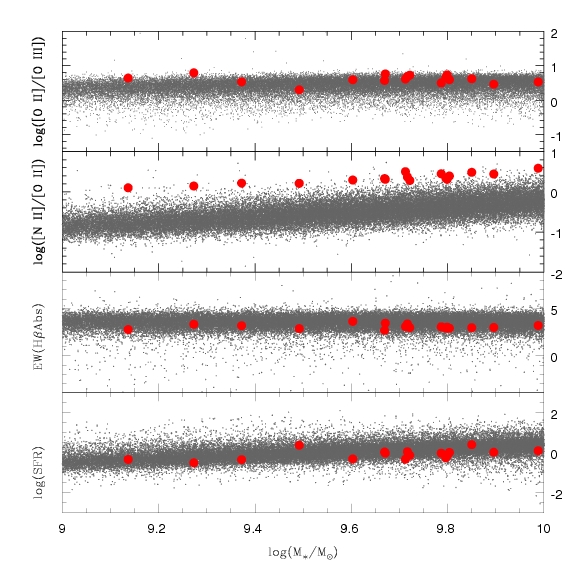}
      \caption{Properties of SDSS star-forming galaxies (gray dots) plotted 
      against stellar mass ($9 \le$ log(M$_{\star}$/M$_{\odot}) \le 10$).
      In comparison, the P08 ``main" sample (the 17 objects with measurable [\ion{O}{2}]) are plotted in red. 
      The top panel shows that in contrast to the ``very low mass" sample, the ``main" sample has normal ionization levels.
      Extremely high N values for the ``main" sample relative to the SDSS average are shown in the second panel.
      The bottom two panels show that these objects have H$\beta$ absorption equivalent widths and SFRs congruent with the SDSS sample. 
      As all objects in the P08 samples exhibit elevated N/O ratios, we suggest N enrichment
      is the prominent factor influencing these objects.}
    \label{fig:10}
  \end{center}
\end{figure}


\subsection{The True Nature of the High Oxygen Abundance Outliers}\label{sec:nature}

Although it appears that the ``very low mass", high oxygen abundance 
galaxies from P08 may not be true outliers from the M-Z relation, 
understanding how these galaxies produced spectra that lie far 
from the median relationships defined by the SDSS galaxies in 
various diagnostic diagrams remains an interesting question.  
Our MMT observations reveal that these spectra are characterized by 
(1) low excitation (low $\lambda$5007/$\lambda$3727 ratios; log(O$_{32}) < -0.4$, see Figures~\ref{fig:8} and ~\ref{fig:9}), 
(2) relatively high N/O abundance ratios for their stellar masses (see Figure~\ref{fig:6}), and 
(3) relatively deep underlying stellar Balmer absorption (ranging from 4 to 8 \AA ).
All three characteristics can be explained by a previous, 
but recent burst of star formation.  
Essentially, these properties are all expected $\sim$ 5 Myr or longer 
after a starburst characterized as a ``Wolf-Rayet galaxy.''

\citet{kunth81} demonstrated that the presence of strong W-R features 
in the spectrum of a star forming region are indicative of an intense 
episode of star formation lasting a relatively short 
(on the order of 10$^6$ years) duration.  
For an instantaneous starburst, these W-R stars appear and disappear 
during the interval of 3 to 6 Myr \citep{leitherer95}.
Afterward, the most massive O stars have all evolved, and the less 
massive O and B stars produce a significantly softer spectrum resulting 
in lower excitation nebulae.
The starburst models of \citet{gonzalez99} show that the equivalent width 
of underlying stellar H$\beta$ absorption increases from roughly 2.7 to 4.5 
{\AA} as an instantaneous starburst of solar metallicity ages from 0 to 10 Myr, 
and continues on to nearly 9 {\AA} at an age of 100 Myr.
On this basis, the MPA-JHU estimates of the Balmer absorption are 
consistent with \ion{H}{2} region ages greater than 4 Myr (3.4 \AA), 
while our estimates (larger due to aperture differences), 
suggest a lower limit of 7 Myr (5.8 \AA).
Finally, during the W-R stage, it is possible for stars to release a large 
amount of N back into the local ISM 
\citep[see, e.g.,][and references therein]{esteban92,kobulnicky97,lse10}, 
producing inflated N/O ratios such as is seen in the four MMT spectra. 


\subsection{P08 ``Main" Sample}\label{sec:MS}

Since all four of our objects, which were picked at random, have roughly identical spectral characteristics, 
we assume that the majority of the ``very low mass" sample of P08 are probably similar.  
The ``main" sample of P08 was defined by different restrictions, including a redshift 
cut to ensure the coverage of both [\ion{O}{2}] and [\ion{O}{3}] in the sample spectra 
(for 17 of the 24 ``main" sample objects).
We obtained the SDSS emission line strengths for the ``main" sample from emission 
line analysis by the MPA-JHU collaboration.
From these data we found the ``main" sample galaxies (those with measured 
[\ion{O}{2}] $\lambda3727$) have normal $\lambda3727/\lambda5007$ ratios for their masses.
In Figure~\ref{fig:10} we show some the properties of the ``main" sample relative to the SDSS 
parent sample, highlighting the normal ionization strengths in the top panel, followed by average
H$\beta$ absorption equivalent widths and SFRs in the bottom two panels.
The ``main" and `very low mass" samples are different in that the ``main" sample galaxies appear
to have normal excitations. 
However, more importantly, in Figure~\ref{fig:10} we have also plotted N/O ratio versus stellar 
mass for the P08 ``main" sample, showing these galaxies all have very high N/O values.
Similar to the four galaxies in our sample, the P08 ``main" sample data 
suggest that they have also experienced recent N enrichment. 
Since high N/O values tend to bias strong line methods towards high abundances, 
and the metallicity offsets for the ``main" sample are smaller than for the 
``very low mass" sample, this bias could easily account for most of this deviation.

All objects of the P08 parent sample display the characteristic nitrogen enhancement 
expected for a galaxy after passing through the ``Wolf-Rayet galaxy" phase.
P08 suggest that these objects are ``transitional objects running out of fuel": 
i.e., the high metallicity results because there is only a small amount of ISM to enrich. 
Our overall conclusion is that the high-metallicity outliers are special, not due to unusually high
oxygen abundances for their mass or rate of star formation, but, rather, due to an
enrichment of nitrogen relative to oxygen.
This nitrogen enrichment may be due to being observed during a rare and short lived evolutionary phase. 


\section{CONCLUSION}
We have investigated four of the 41 luminosity-metallicity outliers from the 
low-mass, high oxygen galaxy sample reported by \citet{peeples08}. 
From new spectral observations taken at the MMT that include the [\ion{O}{2}] 
$\lambda$3727 line,  three of four galaxies were found to have relatively low 
excitation (low $\lambda$5007/$\lambda$3727 ratios;  log(O$_{32}) < -0.4$) 
and, more importantly, all four exhibited high N/O values.  
Each of these characteristics can lead to overestimates of the oxygen abundance 
if standard strong line calibrations are used; furthermore, these two characteristics 
combined can produce the significant metallicity deviations seen for the P08 sample.  
By comparing our four spectra to the photoionization models of \citet{stasinska96}
 (which include the effects of aging of the exciting stars), and the empirical calibrations
 of PMC09 (which correct for variations in the N/O ratio), we found that 
significantly lower oxygen abundances are favored.  
The \citet{stasinska96} oxygen abundances are in line with those expected from 
the luminosities and stellar masses of the galaxies, while the PMC09 
estimates shift oxygen abundances in the right direction.
While the ``very low mass" sample displays low star formation rates,
the normal star formation within the ``main" sample rules out exhaustive star formation as 
the variable responsible for the unusual sample spectra. 
With these conclusions, we propose that these galaxies are best described as nitrogen 
enriched, which may result from having recently passed through the ``Wolf-Rayet galaxy'' phase.


\acknowledgements
We acknowledge Liese van Zee for several helpful discussions
and the anonymous referee for constructive comments and suggestions that greatly
improved the analysis and clarified the text.
DAB is grateful for support from a Penrose Fellowship and a NASA Space 
Grant Fellowship from the University of Minnesota.
EDS is grateful for partial support from the University of Minnesota.

Observations reported here were obtained at the MMT Observatory, a joint
facility of the Smithsonian Institution and the University of Arizona.
MMT observations were obtained as part of the University of Minnesota's 
guaranteed time on Steward Observatory facilities through membership in 
the Research Corporation and it's support for the Large Binocular Telescope.
Funding for the Sloan Digital Sky Survey (SDSS) has been provided by the 
Alfred P.\ Sloan Foundation, the Participating Institutions, the National 
Aeronautics and Space Administration, the National Science Foundation, the U.S. 
Department of Energy, the Japanese Monbukagakusho, and the Max Planck Society. 
The SDSS Web site is http://www.sdss.org/.
This research has made use of NASA's Astrophysics Data System
Bibliographic Services and the NASA/IPAC Extragalactic Database
(NED), which is operated by the Jet Propulsion Laboratory, California
Institute of Technology, under contract with the National Aeronautics
and Space Administration.


\newpage 


\end{document}